\documentclass[12pt]{article}

\usepackage[utf8]{inputenc}
\usepackage[vcentermath]{youngtab}
\usepackage{color}
\usepackage{hyperref}
\usepackage[centertags]{amsmath}
\usepackage{amsfonts} 
\usepackage{amssymb}
\usepackage{amsthm}
\numberwithin{equation}{section} 
\usepackage{graphicx}
\usepackage[compat=1.1.0]{tikz-feynman}
\usepackage{contour}
\definecolor{light-gray}{gray}{0.95}
\usepackage{subcaption}

\begin{document}
\begin{titlepage}
\hfill \hbox{NORDITA-2018-052}
\vskip 0.1cm
\hfill \hbox{QMUL-PH-18-12}
\vskip 1.0cm
\begin{flushright}
\end{flushright}
\vskip 1.0cm
\begin{center}
  {\Large \bf 
    The Subleading Eikonal in Supergravity Theories}
   \vskip 1.0cm {\large Arnau Koemans Collado$^{a}$, Paolo Di Vecchia$^{b, c}$,
Rodolfo Russo$^{a}$, \\ Steven Thomas$^{a}$ } \\[0.7cm]
{\it $^a$ Centre for Research in String Theory, School of Physics and Astronomy \\ Queen Mary University of London, Mile End Road, E1 4NS London, United Kingdom}\\
{\it $^b$ The Niels Bohr Institute, University of Copenhagen, Blegdamsvej 17, \\
DK-2100 Copenhagen, Denmark}\\
{\it $^c$ Nordita, KTH Royal Institute of Technology and Stockholm University, \\Roslagstullsbacken 23, SE-10691 Stockholm, Sweden}\\
\end{center}

\begin{abstract}
In this paper we study the subleading contributions to eikonal scattering in (super)gravity theories with particular emphasis on the role of both elastic and inelastic scattering processes. For concreteness we focus on the scattering of various massless particles off a stack of D$p$-branes in type II supergravity in the limit of large impact parameter $b$. We analyse the relevant field theory Feynman diagrams which naturally give rise to both elastic and inelastic processes. We show that in the case analysed the leading and subleading eikonal only depend on elastic processes, while inelastic processes are captured by a pre-factor multiplying the exponentiated leading and subleading eikonal phase. In addition to the traditional Feynman diagram computations mentioned above, we also present a novel method for computing the amplitudes contributing to the leading and subleading eikonal phases, which, in the large $b$ limit, only involves knowledge of the onshell three and four-point vertices. The two methods are shown to give the same results. Furthermore we derive these results in yet another way, by computing various one-point amplitudes which allow us to extract the classical solution of the gravitational back reaction of the target D$p$-branes. Finally we show how our expressions for the leading and subleading eikonal agree with the calculation of the metric and corresponding deflection angle for massless states moving along geodesics in the relevant curved geometry.
\end{abstract}

\end{titlepage}

\section{Introduction }
\label{intro}

In the Regge high energy limit the $2\to 2$ scattering process is dominated by the contributions of the highest spin states in the theory~\cite{tHooft:1987vrq, Amati:1987wq, Muzinich:1987in}. So, in a gravitational theory that reduces to (super)gravity at large distances, this scattering is dominated at large values of the impact parameter by ladder diagrams involving the exchange of gravitons between the external states. The leading energy contributions of this class of diagrams resums into an exponential; this is the so-called eikonal phase and is directly related to the classical quantities characterising this scattering such as the deflection angle or the time delay.

While this picture applies to any weakly coupled gravitational theory, new features arise when one goes beyond two derivative gravity. For instance, in string theory the eikonal phase is promoted to an eikonal operator; since we are now dealing with objects that have a characteristic length, in certain regimes tidal forces~\cite{Giddings:2006vu,D'Appollonio:2013hja} can become important and excite the incoming state to different final states so as to produce an inelastic transition. At the leading order in the high energy, large impact parameter expansion, this stringy eikonal operator is obtained~\cite{Amati:1987wq,Amati:1987uf,Amati:1988tn,D'Appollonio:2010ae} from the standard eikonal phase, written in terms of the impact parameter $b$, simply via a shift $b\to b+ \hat{X}$, where $\hat{X}$ contains the bosonic string oscillation modes. A non-trivial eikonal operator also appears in the context of a gravitational effective field theory with higher derivative terms that modify the onshell 3-graviton vertex~\cite{Camanho:2014apa}. If the scale $\ell_{hd}$ at which the higher derivative corrections become important is much bigger than the Planck scale $\ell_P$, then,  by resumming the leading energy behaviour of the ladder diagrams as mentioned above, it is possible to use the effective field theory description to derive an eikonal operator also valid at scales $b \sim \ell_{hd}\gg \ell_P$. Again from this result it is possible to derive classical quantities, such as the time delay, that are now obtained from the eigenvalues of the eikonal operator. Generically when $b \sim \ell_{hd}$ the time delay for some scattering process calculated in the effective field theory becomes negative. This causality violation most likely signals a breakdown of the effective field theory approach and in fact is absent when the same process is studied in a full string theory setup~\cite{Camanho:2014apa,DAppollonio:2015fly}.

Since the appearance of inelastic processes in the leading eikonal approximation is the signal of novel physical phenomena such as those mentioned above, it is interesting to see whether there are new features of this type in the {\em subleading} eikonal, which captures the first corrections in the large impact parameter expansion for the same $2 \to 2$ scattering. The aim of this paper is to provide an explicit algorithm that allows us to derive this subleading eikonal from the knowledge of the amplitudes contributing to the scattering process under consideration. In the literature there are several explicit calculations of the subleading eikonal in various gravitational field theories in the two derivative approximation, see for instance~\cite{D'Appollonio:2010ae, Giddings:2010pp,Akhoury:2013yua,Melville:2013qca,Bjerrum-Bohr:2014zsa,Bjerrum-Bohr:2016hpa,Luna:2016idw}. However in these studies the process was {\em assumed} to be elastic to start with, while here we wish to spell out the conditions under which this is the case. Hopefully this will also provide a step towards a full understanding of the subleading eikonal operator at the string level~\cite{Amati:1990xe}. Another goal of our analysis is to highlight that both the leading and the subleading eikonal depend on onshell data. The leading eikonal follows from the spectrum of the highest spin states and the onshell three-point functions, while in the case of the subleading eikonal some further information is necessary as new states in the spectrum may become relevant and the onshell four-point functions provide a non-trivial contribution. 

For the sake of concreteness we cast our analysis in the setup of type II supergravities focusing on the scattering of massless states off a stack of $N$ D$p$-branes~\cite{D'Appollonio:2010ae}, but the same approach can be applied in general to capture the subleading contributions of the large impact parameter scattering in any gravitational theory. In the limit where the mass (density) $N T_p / \kappa_D$ of the target D$p$-branes is large and the gravitational constant  $\kappa_D$ is small, with $N T_p \kappa_D$ fixed,  the process describes the scattering in a classical potential given by the gravitational backreaction of the target. In this case the eikonal phase is directly related (by taking its derivative with respect to the impact parameter) to the deflection angle of a geodesic in a known background. When considering the scattering of a dilaton in the maximally supersymmetric case, there is perfect agreement for the deflection angle between the classical geodesic and the amplitude calculations including the first subleading order~\cite{D'Appollonio:2010ae}. However, in the Feynman diagram approach there are inelastic processes, where a dilaton is transformed into a Ramond-Ramond (RR) field, at the same order in energy as the elastic terms contributing to the subleading eikonal (see section~\ref{ineldilRRallE}). Thus it is natural to ask what the role of these inelastic contributions is and why they should not contribute to the classical eikonal even if they grow with the energy of the scattering process. We will see in section~\ref{HElimit} that these contributions arise from the interplay of the {\em leading} eikonal and the inelastic part of the tree-level S-matrix. One should subtract these types of contributions from the expression for the amplitude in order to isolate the terms that exponentiate to provide the classical eikonal. In the example of the dilaton scattering off a stack of D$p$-branes analysed in detail here, this subtraction cancels completely the contribution of the inelastic processes and one recovers for the subleading eikonal the result found in~\cite{D'Appollonio:2010ae}. In more general setups or at further subleading orders this procedure may be relevant for isolating the terms that are exponentiated even in the elastic channel and thus providing a precise algorithm for extracting the classical contribution (the eikonal) from a Feynman diagram calculation may assist in analysing them. It will be interesting to study this problem for the centre of mass scattering of two semiclassical objects, since this result can provide valuable information for the one body effective action~\cite{Buonanno:1998gg} in the post-Minkowskian approximation at the subleading orders, which is used in the analysis of gravitational waves~\cite{Damour:2016gwp,Damour:2017zjx}.

The paper is structured as follows. In section~\ref{1braneamps} we briefly review the kinematics of the process under study and provide the results for the tree-level amplitudes describing the elastic dilaton to dilaton and the inelastic dilaton to RR scatterings. In section~\ref{2braneamps} we study the one-loop diagrams that contribute to the same processes. We perform the calculation in two ways; one is the traditional approach of using Feynman rules, while in a second approach we provide a prescription for treating onshell bulk amplitudes as effective vertices and gluing them to the D$p$-branes. We check that these two approaches provide the same classical eikonal since they agree at the level of the amplitudes except for possible contributions that are localised on the D$p$-branes (i.e. terms that are proportional to a delta function in the impact parameter space). In section~\ref{HElimit} we study the Regge high energy limit of the amplitudes we derived and, as mentioned above, provide a prescription to extract the classical eikonal at subleading orders from the amplitude. In section~\ref{lntoeikonal} we rederive the same diagrams analysed in section~\ref{2braneamps} in a slightly different way, which allows us to extract the classical solution representing the gravitational backreaction of the target D$p$-branes. In this section we also compare the eikonal with the appropriate classical deflection angle. In all our calculations the contributions of the different fields are separated, thus it is straightforward to focus just on the graviton exchanges and obtain both the metric and the deflection angle for pure Einstein gravity which agrees with the results in the literature (see~\cite{Bjerrum-Bohr:2016hpa,Akhoury:2013yua,Luna:2016idw} and references therein). In section~\ref{sec:discussion} we present our conclusions and discuss some possible applications of our approach.

\section{Scattering in the Born Approximation} \label{1braneamps}

In the Born approximation the interaction between a perturbative state and a stack of D$p$-branes is described by a tree-level diagram with two external states~\cite{Klebanov:1995ni,Garousi:1996ad,Hashimoto:1996bf}. In the limit where the distance between the D$p$-branes and the external states is large, this interaction is captured by a tree-level Feynman diagram with the exchange of a single massless state between the D$p$-branes and a bulk three-point vertex. In this section, we briefly summarise the kinematics of this interaction and then discuss its large energy behaviour. The leading term in this limit is dominated by the exchange of the particles with the highest spin; here we focus mainly on the field-theory limit of the full string setup and so the highest spin state is the graviton. This leading term is elastic and universal, {\em i.e.} the polarisation of the in and the out states are identical and, the result depends only on the momentum exchanged and the energy density of the D$p$-brane target. 

In this section we are also interested in the first subleading correction in the large energy limit. As expected, this contribution depends on the exchange of lower spin states, such as the Ramond-Ramond forms in supergravity. This means that the result depends on other features (besides the energy density) of the target D$p$-brane, such as its charge density or its angular momentum. In general at this order, the transition is not elastic and so displays a non-trivial Lorentz structure. As mentioned in the introduction, this result will be important in defining the eikonal limit beyond the leading order in the large distance limit.

\subsection{Kinematics} \label{kinematics}

We can write the momenta of the two massless external particles scattering off a stack of D$p$-branes as follows, 
\begin{equation}
k_1=(E,\ldots,E) \qquad k_2=(-E,\ldots,\mathbf{q}, -E + q_{D-1}) \text{ ,}
\label{k1k2}
\end{equation}
where the dots are over the $p$ spatial components along the D$p$-brane in $k_2$, $\mathbf{q}$ denotes the $D-p-2$ spatial components transverse to the direction of the incoming particle of momentum exchange vector $q=k_1+k_2$ and $q_{D-1}$ is the last component of $q$. Note that in the Regge limit 
\begin{equation}
  \label{eq:ReggeLd}
  q^2 \ll E^2,
\end{equation}
and $q_{D-1}$ is of order $E^{-1}$. Writing out the explicit kinematics as above we can see that $(k_1)_{\parallel}^2=(k_2)_{\parallel}^2=-E^2$ and $(k_1 \cdot k_2)_{\parallel}=E^2$.  Throughout this paper we will be using the following definitions for the Mandelstam variables, $s=-2 k_1 \cdot k_2$, $u=-2 k_1 \cdot k_3$ and $t=-2 k_1 \cdot k_4$.

\subsection{Elastic and Inelastic Diagrams} \label{inelex1}

The elastic scattering of a dilaton with a graviton being exchanged with the D-branes can easily be calculated in supergravity by using the Feynman rules in appendix~\ref{appFeynRules} 
\begin{equation}
  \label{eq:dil-dil0}
  A^{\rm dd}_1 = i (2\pi)^{p+1} \delta^{p+1}(k_1+k_2)\; {\cal A}^{\rm dd}_1\,,~~\mbox{where}~~ {\cal A}^{\rm dd}_1= \frac{2 NT_p \kappa_D E^2}{q^2}~,
\end{equation}
where $N$ is the number of D-branes in the stack. Notice that the result does not depend on the dimensionality $p$ of the D-branes. In the limit~\eqref{eq:ReggeLd}, the leading energy contribution of any elastic scattering is still described by~\eqref{eq:dil-dil0} multiplied by a kinematic factor forcing the polarisation of the ingoing and outgoing polarisation to be the same (for instance, $\epsilon_{1 \mu}^{~~\nu} \epsilon_{2 \nu}^{~~\mu}$ in the graviton-graviton case). For general states there are subleading energy corrections to this formula, but they start at order $E^0$. 

In the inelastic case, in contrast, it is possible to have order $E$ contributions. As an example, let us start from the amplitude where the incoming particle is a dilaton and the outgoing one is an RR state. Again the first amplitude contributing to this process can be derived by using the Feynman rules in appendix~\ref{appFeynRules} 
\begin{equation}
  \label{eq:dil-RR0}
   A^{\rm dR}_1 = i (2\pi)^{p+1} \delta^{p+1}(k_1+k_2)\ \; {\cal A}^{\rm dR}_1\,,~~\mbox{where}~~{\cal A}^{\rm dR}_1= \frac{2 a(D) N T_p \kappa_D E \, q^\mu C_{\mu 1\ldots p}}{q^2}~,
\end{equation}
where $C_{\mu_1 \ldots \mu_{p+1}}$ is the polarisation of the RR potential describing the second external state and $a(D)$ is defined in appendix \ref{appFeynRules}; in 10D type II supergravity we find $a(D=10)=\frac{p-3}{2}$.

Notice that it is possible to derive the same results by using a different approach that uses only on onshell data. The idea is simply to start from an onshell 3-particle vertex in the bulk\footnote{Strictly speaking the onshell vertices between three massless states often vanish in Minkowski space; as usual, one can define a non-trivial three-point vertex by analytic continuation on the momenta or equivalently by thinking of changing the spacetime signature.} instead of using the full Feynman rules. As an example, consider the vertex with two dilatons and one graviton~\eqref{ddgV}: on shell we can ignore the term proportional to $k_1 \cdot k_2 = (k_1+k_2)^2/2=q^2/2$, where $q$ is the momentum of the graviton. When this effective vertex is used in a diagram, we exploit the condition $q^2=0$ to simplify the numerator of the momentum space amplitude. Terms proportional to $q^2$ appearing in the standard Feynman diagram calculation would produce contributions localised on the D-branes as they cancel the pole of the massless propagator $1/q^2$, so we can ignore them for our purposes. Indeed, by using the onshell two dilatons and one graviton vertex, the standard propagator~\eqref{eq:dedop}, multiplying by $-T_p \eta_{\parallel}^{\rho \sigma}$ for the boundary coupling and imposing the onshell conditions in the numerator obtained in this way, one can easily reproduce~\eqref{eq:dil-dil0} up to terms that do not depend on $q$ and so are localised on the D-branes in the impact parameter space (after performing the Fourier transform~\eqref{ttoips}). 

We conclude by mentioning that it is possible to write the amplitudes above including all string theory corrections simply by implementing the following change to the expression ${\cal A}_1$ above
\begin{equation}
  \label{eq:scD}
  T_p \kappa_D \to T_p \kappa_D\; \frac{\Gamma\left(1-\alpha' E^2\right) \Gamma\left(1+ \frac{\alpha' q^2}{4}\right)}{\Gamma\left(1-\alpha' E^2 + \frac{\alpha' q^2}{4} \right)}\sim  T_p \kappa_D\; \Gamma\left(1 + \frac{\alpha'q^2}{4}\right) e^{i \pi \frac{\alpha' q^2}{4}} (\alpha' E^2)^{1-\frac{ \alpha' q^2}{4}}\;,
\end{equation}
where in the final step we have written the result explicitly in the Regge limit.

\section{Double Exchange Scattering} \label{2braneamps}

\begin{figure}[h]
  \centering
  \begin{tikzpicture}[scale=1.5]
    \begin{feynman}
      \vertex[blob, minimum size=1.5cm] (m) at ( 0, 0) {\contour{white}{}};
      \vertex (a) at (-1,-2) {};
      \vertex (b) at ( 1,-2) {};
      \vertex (c) at (-2, 0) {};
      \vertex (d) at ( 2, 0) {};
	  \draw[fill=light-gray] (-1,-2) ellipse (0.5cm and 0.25cm);
	  \draw[fill=light-gray] (1,-2) ellipse (0.5cm and 0.25cm);       
      \diagram* {
      (a) -- [fermion,edge label=$k_3$] (m),
      (b) -- [fermion,edge label=$k_4$,swap] (m),
      (c) -- [fermion,edge label=$k_1$] (m),
      (d) -- [fermion,edge label=$k_2$, swap] (m),
      };
    \end{feynman}
  \end{tikzpicture}
  \caption{A schematic diagram showing our procedure for calculating effective one-loop amplitudes. The circular blob represents the four-point effective vertex and the two oval blobs represent the D-branes. The four-point vertex is sewed with the D-branes by using the appropriate propagator and boundary coupling. \label{fig:1lc}}
\end{figure}
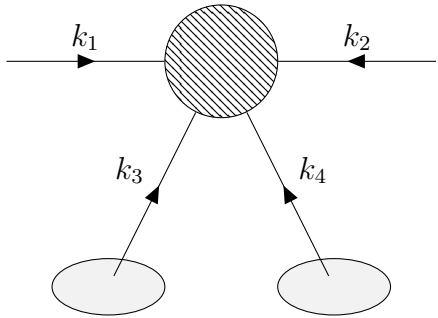

In this section we use the onshell approach mentioned in the previous section to calculate the  amplitudes with a double exchange of particles between the probe and the D-branes. As before we are interested in the classical limit where the gravitational constant is small; $\kappa_D \to 0$, with $N T_p \kappa_D$ fixed. The general idea is that we can use the bulk four-point amplitudes $\mathcal{A}_{bulk}$ as effective vertices and sew them with the relevant propagator to the D-branes, so as to construct diagrams such as the one sketched schematically  in figure~\ref{fig:1lc}.

Again this procedure requires an offshell extension of the bulk four-point vertex, but, as we argue below, the ambiguity related to this step is irrelevant for the large distance (small $q$) scattering. Thus after sewing the D-brane boundary couplings to the relevant external legs of the onshell effective vertex, schematically, we can write the double exchange amplitude as,
\begin{equation}
\mathcal{A}_{2} = \int \frac{d^{\perp}k_i}{(2 \pi)^{\perp}} \frac{d^{\perp}k_j}{(2 \pi)^{\perp}} \left( \frac{1}{2} [B_i] \frac{[G_i]}{k_i^2}\; [B_j] \frac{[G_j]}{k_j^2} \; \delta^{\perp}{(k_i + k_j - q)} \mathcal{A}_{bulk}(k_1,\ldots,k_4)\right) \text{ ,}\label{attachbtoamp}
\end{equation}
where $\perp=D-p-1$ is the number of directions transverse to the D-branes and the overall factor of $1/2$ is a symmetry factor due to the two identical sources. Here we have ``attached" the $i^{th}$ and $j^{th}$ external leg by using the boundary couplings $[B_i]$ and the standard propagators $[G_i]$, see appendix~\ref{appFeynRules}. Terms proportional to $k_i^2$ or $k_j^2$ in $\mathcal{A}_{bulk}$ are absent in the onshell result and would kill one of the propagators attached to the D-branes. We then have a variation of the cancelled propagator argument discussed after~\eqref{eq:dil-RR0}; terms without one of the propagators attached to the D-branes either yield integrals without scale and so can be set to zero in dimensional regularisation or can only produce contributions that are independent of $q$ and so are localised on the D-branes.

\begin{figure}[h]
  \begin{subfigure}[t]{0.3\textwidth}
    \centering
    \begin{tikzpicture}
	    \begin{feynman}
			\vertex (a) at (-1,-2) {};
			\vertex (b) at ( 1,-2) {};
			\vertex (c) at (-2.5, 0) {};
			\vertex (d) at ( 2.5, 0) {};
			\vertex[circle,inner sep=0pt,minimum size=0pt] (e) at (-1, 0) {};
			\vertex[circle,inner sep=0pt,minimum size=0pt] (f) at (1, 0) {};
			\draw[fill=light-gray] (-1,-2) ellipse (0.5cm and 0.25cm);
			\draw[fill=light-gray] (1,-2) ellipse (0.5cm and 0.25cm);       
			\diagram* {
			(c) -- [fermion,edge label=$k_1$] (e) -- [scalar] (f) -- [anti fermion,edge label=$k_2$] (d),
			(a) -- [scalar] (e),
			(b) -- [scalar] (f),
			};
	    \end{feynman}
    \end{tikzpicture}
    \caption{}
    \label{fig:2a}
  \end{subfigure}
  \quad
  \begin{subfigure}[t]{0.3\textwidth}
    \centering
    \begin{tikzpicture}
		\begin{feynman}
			\vertex (a) at (-1,-2) {};
			\vertex (b) at ( 1,-2) {};
			\vertex (c) at (-2.5, 0) {};
			\vertex (d) at ( 2.5, 0) {};
			\vertex[circle,inner sep=0pt,minimum size=0pt] (e) at (0, -1) {};
			\vertex[circle,inner sep=0pt,minimum size=0pt] (m) at (0, 0) {};
			\draw[fill=light-gray] (-1,-2) ellipse (0.5cm and 0.25cm);
			\draw[fill=light-gray] (1,-2) ellipse (0.5cm and 0.25cm);       
			\diagram* {
			(c) -- [fermion,edge label=$k_1$] (m) -- [anti fermion,edge label=$k_2$] (d),
			(m) -- [boson] (e),
			(a) -- [scalar] (e),
			(b) -- [scalar] (e),
			};
		\end{feynman}
    \end{tikzpicture}
    \caption{}
    \label{fig:2b}
  \end{subfigure}
  \quad
  \begin{subfigure}[t]{0.3\textwidth}
    \centering
    \begin{tikzpicture}
    	\begin{feynman}
			\vertex (a) at (-1,-2) {};
			\vertex (b) at ( 1,-2) {};
			\vertex (c) at (-2.5, 0) {};
			\vertex (d) at ( 2.5, 0) {};
			\vertex[circle,inner sep=0pt,minimum size=0pt] (e) at (0, -1) {};
			\vertex[circle,inner sep=0pt,minimum size=0pt] (m) at (0, 0) {};
			\draw[fill=light-gray] (-1,-2) ellipse (0.5cm and 0.25cm);
			\draw[fill=light-gray] (1,-2) ellipse (0.5cm and 0.25cm);       
			\diagram* {
			(c) -- [fermion,edge label=$k_1$] (m) -- [anti fermion,edge label=$k_2$] (d),
			(a) -- [scalar] (m),
			(b) -- [scalar] (m),
			};
		\end{feynman}
    \end{tikzpicture}
    \caption{}
    \label{fig:2c}
  \end{subfigure}
  \caption{The various topologies of diagrams that contribute to $\mathcal{A}^{\rm ddRR}$. In \ref{fig:2a} we have the t- and u-channels, in \ref{fig:2b} we have the s-channel diagram and finally in \ref{fig:2c} we have the contact diagram. The solid lines represent dilatons, wavy lines represent gravitons and the dashed lines represent RR fields.}
  \label{fig:2}
\end{figure}
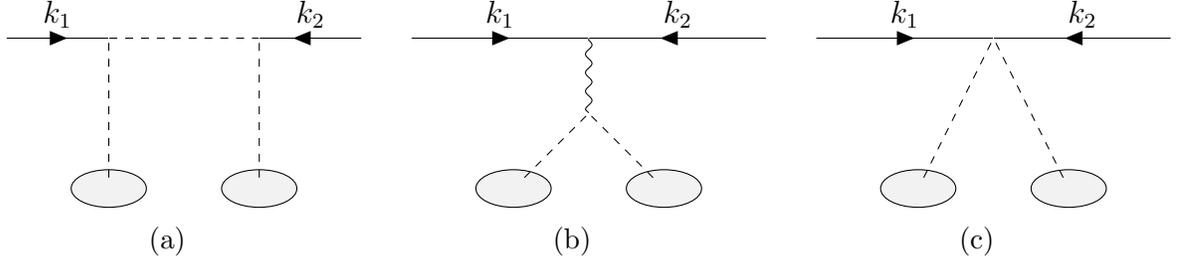

\begin{figure}[h]
  \begin{subfigure}[t]{0.3\textwidth}
    \centering
    \begin{tikzpicture}
	    \begin{feynman}
			\vertex (a) at (-1,-2) {};
			\vertex (b) at ( 1,-2) {};
			\vertex (c) at (-2.5, 0) {};
			\vertex (d) at ( 2.5, 0) {};
			\vertex[circle,inner sep=0pt,minimum size=0pt] (e) at (-1, 0) {};
			\vertex[circle,inner sep=0pt,minimum size=0pt] (f) at (1, 0) {};
			\draw[fill=light-gray] (-1,-2) ellipse (0.5cm and 0.25cm);
			\draw[fill=light-gray] (1,-2) ellipse (0.5cm and 0.25cm);       
			\diagram* {
			(c) -- [fermion,edge label=$k_1$] (e) -- [fermion] (f) -- [anti fermion,edge label=$k_2$] (d),
			(a) -- [boson] (e),
			(b) -- [boson] (f),
			};
	    \end{feynman}
    \end{tikzpicture}
    \caption{}
    \label{fig:3a}
  \end{subfigure}
  \quad
  \begin{subfigure}[t]{0.3\textwidth}
    \centering
    \begin{tikzpicture}
		\begin{feynman}
			\vertex (a) at (-1,-2) {};
			\vertex (b) at ( 1,-2) {};
			\vertex (c) at (-2.5, 0) {};
			\vertex (d) at ( 2.5, 0) {};
			\vertex[circle,inner sep=0pt,minimum size=0pt] (e) at (0, -1) {};
			\vertex[circle,inner sep=0pt,minimum size=0pt] (m) at (0, 0) {};
			\draw[fill=light-gray] (-1,-2) ellipse (0.5cm and 0.25cm);
			\draw[fill=light-gray] (1,-2) ellipse (0.5cm and 0.25cm);       
			\diagram* {
			(c) -- [fermion,edge label=$k_1$] (m) -- [anti fermion,edge label=$k_2$] (d),
			(m) -- [boson] (e),
			(a) -- [boson] (e),
			(b) -- [boson] (e),
			};
		\end{feynman}
    \end{tikzpicture}
    \caption{}
    \label{fig:3b}
  \end{subfigure}
  \quad
  \begin{subfigure}[t]{0.3\textwidth}
    \centering
    \begin{tikzpicture}
    	\begin{feynman}
			\vertex (a) at (-1,-2) {};
			\vertex (b) at ( 1,-2) {};
			\vertex (c) at (-2.5, 0) {};
			\vertex (d) at ( 2.5, 0) {};
			\vertex[circle,inner sep=0pt,minimum size=0pt] (e) at (0, -1) {};
			\vertex[circle,inner sep=0pt,minimum size=0pt] (m) at (0, 0) {};
			\draw[fill=light-gray] (-1,-2) ellipse (0.5cm and 0.25cm);
			\draw[fill=light-gray] (1,-2) ellipse (0.5cm and 0.25cm);       
			\diagram* {
			(c) -- [fermion,edge label=$k_1$] (m) -- [anti fermion,edge label=$k_2$] (d),
			(a) -- [boson] (m),
			(b) -- [boson] (m),
			};
		\end{feynman}
    \end{tikzpicture}
    \caption{}
    \label{fig:3c}
  \end{subfigure}
  \caption{The various topologies of diagrams that contribute to $\mathcal{A}^{\rm ddgg}$. In \ref{fig:3a} we have the t- and u-channels, in \ref{fig:3b} we have the s-channel diagram and finally in \ref{fig:3c} we have the contact diagrams. The solid lines represent dilatons and the wavy lines represent gravitons.}
  \label{fig:3}
\end{figure}
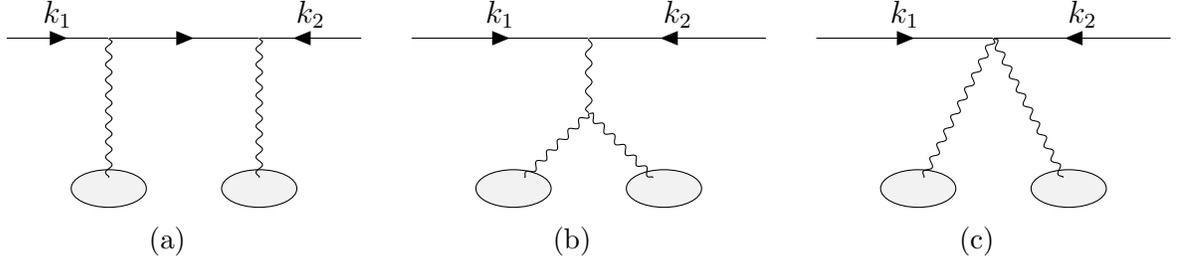

In appendix \ref{integralref} we list the integrals that are relevant for the amplitude in figure~\ref{fig:1lc}. For instance the 3-propagator integral~\eqref{I3full}, relevant for the diagrams in figure~\ref{fig:2a} and figure~\ref{fig:3a}, is
\begin{equation}
  \mathcal{I}_{3}(q_{\perp}) = \int \frac{d^{\perp} k}{(2\pi)^{\perp}}\frac{1}{k^2 (k_1-k)_{\perp}^2 (k+k_2)_{\perp}^2}~.
 \end{equation}
We can see that, when one of the perpendicular propagators is cancelled, the integral can only depend on the quantities $k_1$ (or $k_2$), $\eta_{\mu \nu}$ and $\eta_{\parallel \mu \nu}$. Since these terms do not depend on the exchanged momentum $q=k_1+k_2$, we find that in impact parameter space their contributions are delta functions. The presence of factors proportional to $q$ in the numerator (arising from the vertices) does not spoil the argument, as in this case the impact parameter result is proportional to derivatives of the delta function and so is still localised on the D-branes. Notice that this cancelled propagator argument generalises to the ladder type diagrams with any number of propagators. Another type of integral appearing in the explicit evaluation of~\eqref{attachbtoamp} is (see appendix~\ref{integralref2prop})
\begin{equation}
  \mathcal{I}_{2}(q_{\perp}) = \int \frac{d^{\perp} k}{(2\pi)^{\perp}} \frac{1}{k_{\perp}^2 (k-q)_{\perp}^2}\;.
\end{equation}
In this case, if one of the two propagators in the integrand is cancelled, one obtains an integral without a scale and so again the ambiguities related to the offshell extension of the four-point bulk amplitudes are irrelevant for the calculation we are interested in.

In order to complete the argument and show that using the bulk four-point amplitude in~\eqref{attachbtoamp} is sufficient for our purposes, one should consider also the transverse conditions that are enforced on the onshell vector and graviton fields, such as $k_i^\mu \epsilon^{(i)}_{\mu\nu}=0$ for the case of a graviton. This same issue does not arise when attaching RR fields as we will see in subsection \ref{AttachRR}. We will discuss this point in more detail in the following subsections where we derive~\eqref{attachbtoamp} explicitly for the elastic dilaton-brane scattering amplitudes when the interaction is mediated by RR fields, gravitons and dilatons.

\subsection{Dilaton to Dilaton Elastic Scattering}

We first apply the approach sketched above to the elastic dilaton-brane scattering deriving the full subleading amplitude. From a diagrammatic point of view there are three types of contributions due to the exchange of RR, graviton and dilaton fields between the external particles and the D-branes. We will also compare these results with those obtained from using the supergravity Feynman rules outlined in appendix \ref{appFeynRules}.

\subsubsection{RR Sources} \label{AttachRR}

We start by analysing the RR exchange. By using the four-point two NS-NS (with these states taken to be dilatons), two RR closed string amplitude found in \cite{Bakhtiarizadeh:2013zia} we obtain, in the field theory limit, the following onshell vertex
\begin{eqnarray}
i \mathcal{A}_{bulk}^{\rm ddRR} &=& \frac{i \kappa_D^2}{2}\frac{1}{n!} \frac{2}{s t u} \Bigl[a(D) \, s t u F_{34} + n F_{34}^{\alpha \mu} \Bigl(a(D) \, su k_{2 \alpha} k_{3 \mu}   \nonumber \\
&&  + a(D) \, st k_{2 \mu} k_{4 \alpha} + (2 a^2(D) \, s^2 -8tu)k_{2 \alpha}k_{2 \mu} \Bigr) \Bigr] \nonumber \\
&=& \frac{i \kappa_D^2}{n!} \Bigl[a(D) \, F_{34} + n F_{34}^{\alpha \mu} \Bigl(a(D) \, \frac{1}{t} k_{2 \alpha} k_{3 \mu}   \nonumber \\
&&  + a(D) \, \frac{1}{u} k_{2 \mu}k_{4 \alpha} + \left(2 a^2(D) \, \left(-\frac{1}{t} - \frac{1}{u} \right) - \frac{8}{s} \right) k_{2 \alpha}k_{2 \mu} \Bigr) \Bigr] \text{ ,} \label{STampRR}
\end{eqnarray}
where the various symbols are defined in appendix \ref{appFeynRules} and $n=p+2$. In order to properly attach the D-branes to $\mathcal{A}_{bulk}^{\rm ddRR}$ we need to express $F_{34}^{\alpha \mu}$ as,
\begin{eqnarray}
F_{34}^{\mu \nu} &=& F_3^{\mu \mu_1 \ldots \mu_{n-1}} F_{4 \mu_1 \ldots \mu_{n-1}}^{\nu} \nonumber \\
&=& \left( k_3^{\mu} C^{(3) \mu_1 \ldots \mu_{n-1}} + (-1)^{n-1} k_3^{\mu_1} C^{(3) \mu_2 \ldots \mu_{n-1} \mu} + \ldots \right) \nonumber \\
&& \times \left( k_4^{\nu} C^{(4)}_{\mu_1 \ldots \mu_{n-1}} + (-1)^{n-1} k_{4 \mu_1} C^{(4)}_{\mu_2 \ldots \mu_{n-1}}{}^{\nu} + \ldots \right) \nonumber \\
&=&  k_3^{\mu} k_4^{\nu} C^{(3) \mu_1 \ldots \mu_{n-1}} C^{(4)}_{\mu_1 \ldots \mu_{n-1}} + (n-1) k_3 \cdot k_4 C^{(3) \mu \mu_2 \ldots \mu_{n-1}} C^{(4) \nu}_{\mu_2 \ldots \mu_{n-1}}  \text{ ,}
\end{eqnarray}
where we have used the facts that $k_3$ and $k_4$ only have components perpendicular to the D-branes and $C^{(3)}$ and $C^{(4)}$ only have components parallel to the D-branes, which therefore implies that $k_i \cdot C^{(j)} = 0$. We now take derivatives with respect to the gauge fields to make this an effective vertex to use when we attach the D-branes to the RR fields. We need to also take into account the different sets of labels that the $C$ fields can carry, i.e. $\mu_1\mu_2 \ldots \mu_{n-1} = 01\ldots p$ or $\mu_1\mu_2 \ldots \mu_{n-1} = 12 \ldots p0$,  etc., for which we note there are $(n-1)!$ sets of possible labels for $C^{(3) \mu_1 \ldots \mu_{n-1}} C^{(4)}_{\mu_1 \ldots \mu_{n-1}}$, as there are $(n-1)$ contracted indices, and $(n-2)!$ for $C^{(3) \mu \mu_2 \ldots \mu_{n-1}} C^{(4) \nu}_{\mu_2 \ldots \mu_{n-1}}$. Putting this together allows us to write,

\begin{eqnarray}
F_{34}^{\mu \nu} &=& (n-1)! k_3^{\mu} k_4^{\nu}  + (n-1)(n-2)! k_3 \cdot k_4 \eta_{\parallel}^{\mu \nu} \nonumber \\
&=& (n-1)!(k_3^{\mu} k_4^{\nu}  + k_3 \cdot k_4 \eta_{\parallel}^{\mu \nu}) \text{ .} \label{F12braneattach}
\end{eqnarray}
From the last line above we can also deduce that $F_{34} = n! k_3 \cdot k_4$. Note that these expressions only hold when both RR fields are attached to the D-branes. We can now use \eqref{F12braneattach} as well as $(k_2)_{\parallel}^2=-E^2$ and $(k_2 \cdot k_3)_{\parallel}=(k_2 \cdot k_4)_{\parallel} = 0$ to rewrite the contribution to \eqref{attachbtoamp} due to the onshell vertex \eqref{STampRR} when the RR fields are attached to the D-branes. In this case the integrand of \eqref{attachbtoamp} then reads
\begin{eqnarray}
 \frac{i (N \mu_p\kappa_D)^2}{2} \frac{1}{n!} \frac{(n-1)!}{k_3^2 k_4^2} \Bigl[ 2 a^2(D) n \left( \frac{s}{4} + \frac{s^2}{2 t u}E^2 \right) + n \left( -\frac{2 t u}{s} - 4 E^2 \right) \Bigr] \text{ ,} \label{FullRRSTAmp1}
\end{eqnarray}
where $N$ has been inserted to take into account the $N$ D-branes in the stack. We can write the full answer in terms of the momentum integrals defined in appendix \ref{integralref},
\begin{equation}
i \mathcal{A}_{2}^{\rm ddRR} = i (N T_p\kappa_D)^2 \Bigl[ 2 a^2(D) \, \left( \frac{s}{4} \mathcal{I}_2 + s E^2 \mathcal{I}_3 \right) - \left( \frac{8}{s} k_{1 \mu} k_{2 \nu} \mathcal{I}_{2}^{\mu \nu} + 4 E^2 \mathcal{I}_2 \right) \Bigr] \text{ .} \label{FullRRSTAmp2}
\end{equation}

We want to compare \eqref{FullRRSTAmp2} with the equivalent result arising from performing the same calculation using Feynman diagrams. We can calculate all the relevant onshell Feynman diagrams for this process. The four contributions to the full amplitude are given by,
\begin{eqnarray}
i \mathcal{A}^{\rm ddRR}_{{\rm FT}, u} &=&  [V_{\phi_1 F^{(n)}_{3} C^{(n-1)}}]_{\mu_2 \ldots \mu_n} [V_{\phi_2 F^{(n)}_{4} C^{(n-1)}}]^{\mu_2 \ldots \mu_n} [G_{C^{(n-1)}}] \nonumber \\
&=& \frac{i \kappa_D^2}{(n-1)!} \frac{2 a^2(D)}{(k_1+k_3)^2} F_{34}^{\mu \nu} (k_1+k_3)_{\mu} (k_1+k_3)_{\nu} \label{FTampRR1} \\ 
&& \nonumber \\
i \mathcal{A}^{\rm ddRR}_{{\rm FT}, t} &=&  [V_{\phi_2 F^{(n)}_{3} C^{(n-1)}}]_{\mu_2 \ldots \mu_n} [V_{\phi_1 F^{(n)}_{4} C^{(n-1)}}]^{\mu_2 \ldots \mu_n} [G_{C^{(n-1)}}] \nonumber \\
&=& \frac{i \kappa_D^2}{(n-1)!} \frac{2 a^2(D)}{(k_1+k_4)^2} F_{34}^{\mu \nu} (k_1+k_4)_{\mu} (k_1+k_4)_{\nu}  \label{FTampRR2} \\
&& \nonumber \\
i \mathcal{A}^{\rm ddRR}_{{\rm FT}, s} &=&  [V_{\phi_1 \phi_2 h}]_{\mu \nu} [G_{h}]^{\mu \nu ; \rho \sigma} [V_{F^{(n)}_{3} F^{(n)}_{4} h}]_{\rho \sigma} \nonumber \\
&=& - \frac{2i \kappa_D^2}{n!} \frac{1}{(k_1+k_2)^2} (n F_{34}^{\mu \nu} (k_{1 \mu} k_{2 \nu} + k_{2 \mu} k_{1 \nu}) - k_1 \cdot k_2 F_{34}) \label{FTampRR3} \\
&& \nonumber \\
i \mathcal{A}^{\rm ddRR}_{{\rm FT}, c} &=& [V_{\phi_1 \phi_2 F^{(n)}_{3} F^{(n)}_{4}}] \nonumber \\
&=& - \frac{2 i \kappa_D^2}{n!} a^2(D) F_{34} \text{ ,} \label{FTampRR4}
\end{eqnarray}
where we have neglected to write the various momentum conserving delta functions. For simplicity we have written the expressions above without including the boundary vertex corresponding to the D-branes. In order to obtain the amplitudes with the D-branes attached one needs to multiply the amplitudes above by $[G_{C^{(n-1)}}] [B_{C^{(n-1)}}]$ for every D-brane that is attached.

We now need to sum \eqref{FTampRR1}-\eqref{FTampRR4} and include the factors of $[G_{C^{(n-1)}}] [B_{C^{(n-1)}}]$ we excluded earlier. We also need to use expressions such as, $F_{34}^{\mu \nu} (k_{1 \mu} k_{2 \nu} + k_{2 \mu} k_{1 \nu}) = -2 F_{34}^{\mu \nu} k_{2 \mu}k_{2 \nu} + \frac{k_3 \cdot k_4}{n}F_{34}$, $F^{ \alpha \mu}_{34} k_{2 \alpha} k_{3 \mu} = - \frac{t}{2n} F_{34}$ and $F^{ \alpha \mu}_{34} k_{4 \alpha} k_{2 \mu} = - \frac{u}{2n} F_{34}$, which are straightforward to derive using \eqref{F12braneattach} as a reference. We find that the full amplitude is given by,

\begin{eqnarray}
i \mathcal{A}^{\rm ddRR}_{{\rm FT}} &=& -(i N \mu_p)^2 \frac{1}{2} \int \frac{d^{\perp}k_3}{(2 \pi)^{\perp}} \frac{d^{\perp}k_4}{(2 \pi)^{\perp}} \frac{1}{k_3^2} \frac{1}{k_4^2} \delta^{\perp}{(k_3 + k_4 - q)} \nonumber \\ &&\times \frac{i \kappa_D^2}{2} \frac{2}{(n-1)!} \left( 2 a^2(D) \, \frac{s}{tu} - \frac{8}{s} \right) F_{34}^{\mu \nu} k_{2 \mu}k_{2 \nu} \text{ .}
\end{eqnarray}
Using equation \eqref{F12braneattach} we find,

\begin{eqnarray}
i \mathcal{A}^{\rm ddRR}_{{\rm FT}} = i (N T_p\kappa_D)^2 \Bigl[ 2 a^2(D) \, \left( \frac{s}{4} \mathcal{I}_2 + s E^2 \mathcal{I}_3 \right) - \left( \frac{8}{s} k_{1 \mu} k_{2 \nu} \mathcal{I}_{2}^{\mu \nu} + 4 E^2 \mathcal{I}_2 \right) \Bigr] \text{ .}
\label{FullRRFTAmp}
\end{eqnarray}
Comparing \eqref{FullRRFTAmp} with \eqref{FullRRSTAmp2} we find that we have been able to reproduce the same results we produced using our ``effective bulk vertex" prescription by using traditional supergravity Feynman rules.

\subsubsection{Graviton Sources} \label{AttachGrav}

As we have done in the previous subsection for RR fields, we want to derive the full field theory amplitude for graviton exchange by using the four-point NS-NS closed string amplitude (with two external states taken to be dilatons and two taken to be gravitons) as the effective four-point vertex. When attaching a D-brane sourcing a graviton one replaces the polarisation of the relevant external graviton in $\mathcal{A}_{bulk}^{\rm ddgg}$ as follows
\begin{equation}
\epsilon^{\mu \nu} \to [G_{h}]^{\mu \nu ; \rho \sigma} [B_{h}]_{\rho \sigma} = - N T_p \left(\eta^{\mu \nu}_{\parallel} - \frac{p+1}{D-2} \eta^{\mu \nu} \right) \text{ ,} \label{ebraneattach}
\end{equation}
which is effectively the combination one needs to use in \eqref{attachbtoamp} alongside the bulk vertex in order to obtain the amplitude with the D-branes attached. 

In the case when we sew D-branes that are sourcing gravitons we have the added complication that, as one can see from \eqref{ebraneattach}, the polarisations of the legs we attach the D-branes are neither transverse nor traceless. However the bulk four-point amplitudes we will use as effective four-point vertices in this subsection assume that the external graviton polarisations are traceless and transverse. This implies that by using momentum conservation and the onshell conditions, it is easy to write equivalent onshell vertices that in general yield different results\footnote{For instance by using directly the expression~\eqref{eq:ddggv} one does not obtain~\eqref{STampGG}, as discussed below.} when sewn to the D-branes. Thus we need to add a prescription on what additional properties the effective vertex should have before sewing it to the D-branes. The onshell vertex vanishes for any longitudinal polarisation of any massless particle, i.e. in the case of gravitons it is zero when we substitute $\epsilon_i^{\mu \nu} = \zeta_i^{\mu} k_i^{\nu} + \zeta_i^{\nu} k_i^{\mu}$. Of course when checking this property one needs in general to use momentum conservation and the onshell properties of the remaining external states. However, the momenta of the gravitons glued to the D-branes will appear as integrated variables in the final expression and at that stage it is not always possible to use momentum conservation to write them in terms of the external momenta. Thus we require a further constraint on the onshell bulk effective vertex that can be used to derive a loop diagram: when one of the gravitons that will be glued to the D-branes is longitudinal, the bulk amplitude must vanish whilst not explicitly using momentum conservation in the products $k_i\epsilon_j$ and $\zeta_i k_j$, but only doing so on products between momenta $k_i k_j$ (i.e. only using $s+t+u=0$ in our analysis). In the case of a four-point bulk onshell amplitude, as long as it includes both momenta of the external legs that will be attached to the D-branes in its ``momentum set'' (i.e. the three independent momenta with which the amplitude is expressed), then the condition mentioned above is met.

We start by recalling the field theory limit of the four-point two dilaton, two graviton amplitude \cite{green1988superstring} which we can write as
\begin{eqnarray}\label{eq:ddggv}
i \mathcal{A}_{bulk}^{\rm ddgg} &=& \frac{i \kappa_D^2}{2} \frac{2}{stu} \left( u^2 t^2 \, \epsilon_3^{\mu \nu}\epsilon_{4 \mu \nu} + 4 u^2 \, k_1^{\mu} k_1^{\nu} k_2^{\rho} k_2^{\sigma} \epsilon_{3 \rho \sigma} \epsilon_{4 \mu \nu} - 4 t u^2 \, k_1^{\mu} k_2^{\nu} \epsilon_{3 \nu}{}^{\rho} \epsilon_{4 \mu \rho}  \right. \nonumber \\
&& \left. - 4 t^2 u \, k_1^{\mu} k_2^{\nu} \epsilon_{3 \mu}{}^{\rho} \epsilon_{4 \nu \rho} + 8 u t \, k_1^{\mu} k_1^{\nu} k_2^{\rho} k_2^{\sigma} \epsilon_{3 \mu \rho} \epsilon_{4 \nu \sigma} + 4 t^2 \, k_1^{\mu} k_1^{\nu} k_2^{\rho} k_2^{\sigma} \epsilon_{3 \mu \nu} \epsilon_{4 \rho \sigma} \right) \,.
\end{eqnarray}
In this form the bulk vertex does not satisfy the requirement mentioned above for the two gravitons, but if we take this equation and use momentum conservation to express it using $(k_1, k_3, k_4)$ or $(k_2, k_3, k_4)$, we obtain an expression that can be glued to the D-branes simply by replacing the graviton polarisations with \eqref{ebraneattach}. Then we find for the integrand of \eqref{attachbtoamp}
\begin{eqnarray} \label{STampGG}
\frac{i (N T_p \kappa_D)^2}{4} \frac{1}{k_3^2 k_4^2} \frac{2}{stu} \left(4 E^4 s^2 + 4 E^2 stu + \frac{(D-p-3)(1+p)}{D-2} u^2 t^2 \right) 
\text{ ,} 
\end{eqnarray}
where we have also used the relevant kinematics mentioned in section \ref{kinematics}. By including the appropriate integrals one obtains
\begin{eqnarray}
i \mathcal{A}_{2}^{\rm ddgg} &=& i (N T_p \kappa_D)^2 \left( 4E^4 \mathcal{I}_{3} + \frac{(D-p-3)(1+p)}{D-2}\frac{2}{s} k_{1 \mu} k_{2 \nu} \mathcal{I}_{2}^{\mu \nu} + 2 E^2 \mathcal{I}_{2}  \right) \text{ .} \label{FullGGSTAmp}
\end{eqnarray}

We want to compare \eqref{FullGGSTAmp} with the equivalent result arising from using Feynman diagrams as we have done in the RR case. We first calculate the relevant Feynman diagrams for this process. Note that since we will be attaching the D-branes to the graviton external legs we will not be imposing $k_i \cdot \epsilon_i = 0$ or $\text{Tr}(\epsilon_i)= 0$, i.e. we will keep the gravitons offshell. The four contributions to the full amplitude are given by the following diagrams,
\begin{eqnarray}
i\mathcal{A}^{\rm ddgg}_{{\rm FT}, u} &=& \epsilon_{3 \mu \nu} \epsilon_{4 \rho \sigma} [V_{\phi_1 \phi_2 h}]^{\mu \nu} [V_{\phi_1 \phi_2 h}]^{\rho \sigma} [G_{\phi}] \\
&& \nonumber \\
i\mathcal{A}^{\rm ddgg}_{{\rm FT}, t} &=&  \epsilon_{3 \rho \sigma} \epsilon_{4 \mu \nu} [V_{\phi_1 \phi_2 h}]^{\mu \nu} [V_{\phi_1 \phi_2 h}]^{\rho \sigma} [G_{\phi}] \\
&& \nonumber \\
i\mathcal{A}^{\rm ddgg}_{{\rm FT}, s} &=&  \epsilon_{3 \rho \sigma} \epsilon_{4 \lambda \tau} [V_{\phi_1 \phi_2 h}]^{\gamma \delta} [G_{h}]_{\mu \nu ; \gamma \delta} \left( T^{\mu \nu;\rho\sigma;\lambda \tau} (q,k_3,k_4) + T^{\rho\sigma;\mu \nu;\lambda \tau} (k_3,q,k_4) \right. \\ \nonumber
&+& \left. T^{\mu \nu;\lambda \tau;\rho\sigma} (q,k_4,k_3) + T^{\rho\sigma;\lambda \tau;\mu \nu} (k_3,k_4,q)       + T^{\lambda \tau;\mu \nu;\rho\sigma} (k_4,q,k_3) + T^{\lambda \tau;\rho\sigma;\mu \nu} (k_4,k_3,q) \right)  \\
&& \nonumber \\
i\mathcal{A}^{\rm ddgg}_{{\rm FT}, c} &=& \epsilon_{3 \rho \sigma} \epsilon_{4 \lambda \tau} [V_{\phi \phi' h h'}]^{\rho \sigma \lambda \tau} \text{ ,}
\end{eqnarray}
where we have not explicitly written the resulting Lorentz structure for brevity and $\epsilon_3$, $\epsilon_4$ are the graviton polarisations which need to be replaced with $[G_{h}]^{\mu \nu ; \rho \sigma} [B_{h}]_{\rho \sigma}$. Doing so and using the kinematics outlined in section \ref{kinematics} we have for the u-channel,
\begin{eqnarray}
i \mathcal{A}^{\rm ddgg}_{{\rm FT}, u} &=& -(-i)(-i \kappa_D)^2 (N T_p)^2 4 E^4 \frac{1}{2} \int \frac{d^{\perp}k_3}{(2 \pi)^{\perp}} \frac{d^{\perp}k_4}{(2 \pi)^{\perp}} \frac{1}{k_3^2} \frac{1}{k_4^2} \delta^{\perp}{(k_3 + k_4 - q)} \frac{1}{u} \nonumber \\
&=& i (N T_p \kappa_D)^2 2 E^4 \mathcal{I}_3 \;, \label{GGuFTAmp} 
\end{eqnarray}
with an equivalent contribution for the t-channel. These two diagrams are the only ones that contribute to leading order in energy, $\mathcal{O}(E^3)$, in the full amplitude as we have seen with the result derived from using our effective bulk vertex method. We can now look at the two remaining diagrams which contribute to subleading order in energy, $\mathcal{O}(E^2)$. The s-channel diagram gives,
\begin{eqnarray}
&& i \mathcal{A}^{\rm ddgg}_{{\rm FT}, s} = - (-i \kappa_D)(-2 i \kappa_D) \left( - \frac{i}{2} \right) (N T_p)^2 \frac{1}{2} \int \frac{d^{\perp}k_3}{(2 \pi)^{\perp}} \frac{d^{\perp}k_4}{(2 \pi)^{\perp}} \frac{1}{k_3^2} \frac{1}{k_4^2} \delta^{\perp}{(k_3 + k_4 - q)} \nonumber \\
&& \times \left[ 4 E^2 \frac{(D-2p-4)}{D-2} + \frac{(D-p-3)(1+p)}{D-2} \left( \frac{4}{s}(k_2 \cdot k_3)(k_2 \cdot k_3) + 2 (k_1 \cdot k_3) \right) \right] \text{ .} \label{GGsFTAmp}
\end{eqnarray}
We also have for the contact diagram,
\begin{eqnarray}
i \mathcal{A}^{\rm ddgg}_{{\rm FT}, c} &=& \left( \frac{i \kappa_D^2}{2} \right) (N T_p)^2 \frac{1}{2} \int \frac{d^{\perp}k_3}{(2 \pi)^{\perp}} \frac{d^{\perp}k_4}{(2 \pi)^{\perp}} \frac{1}{k_3^2} \frac{1}{k_4^2} \delta^{\perp}{(k_3 + k_4 - q)} \biggl( \frac{16 E^2 (D-p-3)}{D-2} \nonumber \\
&& + \frac{4(1+p)(D-p-3)}{D-2} (k_1 \cdot k_2) \biggr) \text{ .} \label{GGcFTAmp} 
\end{eqnarray}
Summing the above two contributions yields
\begin{eqnarray}
i \mathcal{A}^{\rm ddgg}_{{\rm FT}, c} + i \mathcal{A}^{\rm ddgg}_{{\rm FT}, s} &=& i (N T_p \kappa_D)^2 \int \frac{d^{\perp}k_3}{(2 \pi)^{\perp}} \frac{d^{\perp}k_4}{(2 \pi)^{\perp}} \frac{1}{k_3^2} \frac{1}{k_4^2} \biggl( 4 E^2 \nonumber \\
&& +  \frac{(D-p-3)(1+p)}{D-2} (k_1 \cdot k_3) (k_2 \cdot k_3) \biggr) \text{ .} \label{GGcsFTAmp}
\end{eqnarray}
We can easily see by summing \eqref{GGuFTAmp} and \eqref{GGcsFTAmp} that we are able to reproduce \eqref{FullGGSTAmp} using the supergravity Feynman rules.

\subsubsection{Dilaton Sources}\label{AttachDil}

Here we calculate the amplitude for elastic dilaton-brane scattering with dilaton exchange by using the four-point dilaton string amplitude as the effective vertex. We have in the field theory limit of the string theory amplitude \cite{Garousi:2012yr},
\begin{eqnarray}
i \mathcal{A}_{bulk}^{\rm dddd} = i \kappa_D^2 \left( \frac{st}{u} + \frac{su}{t} + \frac{ut}{s} \right) \text{ .}
\end{eqnarray}
As before, using our prescription, we include the relevant integrals arising from \eqref{attachbtoamp}. When attaching dilatons the relevant factor, arising from $[G_{\phi}] [B_{\phi}]$, is $-(i N T_p \frac{a(D)}{\sqrt{2}})^2$. Using this we have,
\begin{eqnarray}
i \mathcal{A}_{2}^{\rm dddd} &=& - i \left( N T_p \kappa_D \frac{a(D)}{\sqrt{2}} \right)^2 \left( s k_{2 \mu} \mathcal{I}_{2}^{\mu} + s k_{1 \mu} \mathcal{I}_{2}^{\mu} + \frac{2}{s} k_{1 \mu} k_{2 \nu} \mathcal{I}_{2}^{\mu \nu}  \right) \text{ .}
\end{eqnarray}
It is trivial to compare these results with those found using supergravity Feynman rules and we will not be comparing them explicitly here.

\subsection{Dilaton to RR Inelastic Scattering} \label{ineldilRRallE}

\begin{figure}[h]
  \centering
  \begin{tikzpicture}[scale=1.5]
    \begin{feynman}
      \vertex[blob, minimum size=1.5cm] (m) at ( 0, 0) {\contour{white}{}};
      \vertex (a) at (-1,-2) {};
      \vertex (b) at ( 1,-2) {};
      \vertex (c) at (-2, 0) {};
      \vertex (d) at ( 2, 0) {};
	  \draw[fill=light-gray] (-1,-2) ellipse (0.5cm and 0.25cm);
	  \draw[fill=light-gray] (1,-2) ellipse (0.5cm and 0.25cm);       
      \diagram* {
      (a) -- [boson,edge label=$k_3$] (m),
      (b) -- [scalar,edge label=$k_4$,swap] (m),
      (c) -- [edge label=$k_1$] (m),
      (d) -- [scalar,edge label=$k_2$, swap] (m),
      };
    \end{feynman}
  \end{tikzpicture}
  \caption{A schematic diagram showing our procedure for calculating the effective one-loop amplitude for dilaton to RR inelastic scattering. The circular blob represents the four-point effective vertex $\mathcal{A}_{bulk}^{\rm dRgR}$ and the two oval blobs represent the D-branes. As before the solid lines correspond to dilatons, the wavy lines correspond to gravitons and the dashed lines correspond to RR fields. \label{fig:4}}
\end{figure}
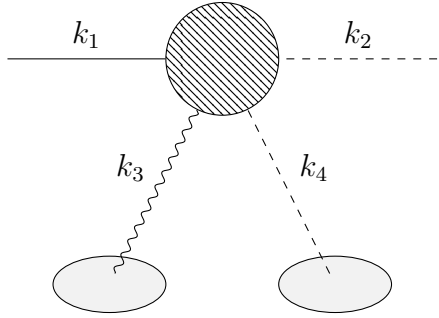

As with the elastic dilaton scattering case that we have considered in the previous subsection we can use equation (47) in \cite{Bakhtiarizadeh:2013zia} for the four-point two NS-NS (with one state taken to be a dilaton and the other a graviton), two RR closed string amplitude as an effective vertex for calculating the amplitude for an inelastic transition from a dilaton to an RR field via the exchange of a graviton and an RR field with the D-branes. The bulk vertex needed is given by,
\begin{eqnarray}
&& i \mathcal{A}_{bulk}^{\rm dRgR} = - \frac{i \kappa_D^2}{\sqrt{2}}\frac{16 a(D)}{n!} \Bigl[ F_{24} \epsilon_3^{\mu \nu} \left( s^2 k_{2 \mu} k_{2 \nu} + t k_{4 \mu} \left( t k_{4 \nu} - 2 s k_{2 \nu} \right) \right)  \nonumber \\
&&  + n u \left( (n-1) u F_{24}^{\alpha \beta \mu \nu} \epsilon_{3 \beta \nu} k_{3 \alpha} k_{3 \mu} - F_{24}^{\alpha \mu} \left(t k_4^{\nu} - s k_2^{\nu} \right)\left( \epsilon_{3 \mu \nu} k_{3 \alpha} - \epsilon_{3 \alpha \nu} k_{3 \mu} \right)      \right) \Bigr] \;, \label{stringdRgRvertex}
\end{eqnarray}
where the labels here correspond as follows; label 1 is associated with the external dilaton, label 2 is associated with the external RR field, label 3 is associated with the internal graviton and label 4 is associated with the internal RR field as shown in figure \ref{fig:4}. After using our prescription and expressing \eqref{stringdRgRvertex} with momentum set $(k_1, k_3, k_4)$ we find that the expression satisfies the condition described at the beginning of section \ref{2braneamps} as required.

By following the same approach discussed in the previous subsection, we find that after attaching the graviton to the D-branes, the integrand of \eqref{attachbtoamp} reads
\begin{eqnarray}
&& \frac{16 i N^2 T_p \mu_p \kappa_D^2}{\sqrt{2}} \frac{1}{k_3^2}  \frac{1}{k_4^2} \frac{a(D)}{n!} \biggl[\frac{s}{ut} F_{24} E^2 + F_{24} \frac{1+p}{D-2} \frac{s}{t} + n F_{24}^{\alpha \beta} \left\lbrace \frac{1}{t} \left(\eta_{\parallel \beta \nu}k_1^{\nu}k_{3 \alpha} \right. \right. \nonumber \\
&& \left. \left. - \eta_{\parallel \alpha \nu}k_1^{\nu}k_{3 \beta} \right)  + \frac{1+p}{D-2} \left( \frac{1}{t} \left(k_{1 \alpha} k_{3 \beta} - k_{1 \beta} k_{3 \alpha} \right) + \frac{u}{ts} \left((n-1)k_{3 \alpha} k_{3 \beta} - k_{4 \alpha}  k_{3 \beta} \right) \right) \right\rbrace \nonumber \\
&&  - \frac{u}{ts} F_{24}^{\alpha \beta \mu \nu} \eta_{\parallel \beta \nu}k_{3 \alpha} k_{3 \mu} \biggr] \;,
\label{RRRRdgrawamp}
\end{eqnarray}
where the first term has been identified in advance as the term with potential to contribute to the leading energy behaviour of the amplitude. In order to attach an RR leg to the D-branes, we need to rewrite the field strengths in terms of the potentials. We have explicitly calculated all the necessary combinations of field strengths that arise in \eqref{RRRRdgrawamp} in appendix \ref{RRidentities}. Using these expressions we can write the integrand in \eqref{attachbtoamp} relevant to the inelastic dilaton to RR amplitude,
\begin{eqnarray}
&& - i (N^2 T_p \mu_p \kappa_D^2  ) \frac{16 a(D)}{\sqrt{2}} \frac{1}{k_3^2 k_4^2} \biggl[ \frac{s}{ut} \left( -E^3 (k_4 \cdot C^{(2)})^{1 \ldots p} - E^2 (k_2 \cdot k_4) C^{(2) 0 \ldots p} \right) - \frac{u}{4} C^{(2) 0 \ldots p} \nonumber \\
&& + \frac{E}{2} (k_4 \cdot C^{(2)})^{1 \ldots p} + \frac{1+p}{D-2} \biggl\lbrace C^{(2) 0 \ldots p} \left( \frac{s}{2t}E^2 + \frac{(3+n)u}{4} + \frac{u^2}{4t} \right) + \frac{s}{2t}E (k_4 \cdot C^{(2)})^{1 \ldots p} \nonumber \\
&& (k_3 \cdot C^{(2)})^{1 \ldots p} E \left( \frac{1}{2} - \frac{u}{2t}n \right) \biggr\rbrace \biggr] \;.
\end{eqnarray}
Inserting the integrals as per our prescription we obtain,
\begin{eqnarray}
\label{eq:dRgRf}
&& i \mathcal{A}_{2}^{\rm dRgR} = - i (N^2 T_p \mu_p \kappa_D^2) \frac{8 a(D)}{\sqrt{2}} \biggl[ - ( q \cdot C^{(2)})^{1 \ldots p} E^3 \mathcal{I}_3 - (q \cdot k_2) C^{(2) 0 \ldots p} E^2 \mathcal{I}_3 + \frac{1}{2} k_{1 \mu} \mathcal{I}_2^{\mu} C^{(2) 0 \ldots p} \nonumber \\
&& + \frac{1}{2}E C_{\mu}^{(2)}{}^{1 \ldots p} \mathcal{I}_2^{\mu} + \frac{1+p}{D-2} \biggl\lbrace C^{(2) 0 \ldots p} \left( -\frac{s}{2} E^2 \mathcal{I}_3 - \frac{(2+n)}{2} k_{1 \mu} \mathcal{I}_2^{\mu} + \frac{s}{4}\mathcal{I}_2 - \frac{s^2}{4}\mathcal{I}_3 \right) \nonumber \\
&& - \frac{s}{4}E ( q \cdot C^{(2)})^{1 \ldots p} \mathcal{I}_3 + \frac{1}{2}E C_{\mu}^{(2)}{}^{1 \ldots p} \mathcal{I}_2^{\mu} + E C_{\mu}^{(2)}{}^{1 \ldots p} \left(\frac{1}{2} \mathcal{I}_2^{\mu} - \frac{s}{4}q^{\mu} \mathcal{I}_3 \right) \biggr\rbrace \biggr] \;,
\end{eqnarray}
where $q=k_1+k_2$ is the momentum exchanged.

\section{The Supergravity Eikonal} \label{HElimit}

In this section we will focus on and derive explicit expressions for the leading and subleading high-energy behaviour of the various amplitudes we considered in section \ref{2braneamps} and analyse their behaviour in the context of the eikonal approximation. 

We can transform any of the amplitudes we have considered into impact parameter space by using,
\begin{equation}\label{ttoips}
\mathcal{A}_{h}(E,b) = \int \frac{\text{d}^{D-p-2} \mathbf{q}}{(2 \pi)^{D-p-2}} e^{i \mathbf{b} \cdot \mathbf{q}} \mathcal{A}_{h}(E,q) \text{ ,}
\end{equation}
where $h$ refers to the number of boundaries of the amplitude (i.e. the number of exchanges with the D-branes). We start by focusing on the elastic case where the leading energy behaviour of the tree-level amplitude, one-loop amplitude and amplitudes with a higher number of boundaries is universal and so does not display any non-trivial Lorentz structure. By summing these contributions, we find that the S-matrix approximates to,
\begin{equation} \label{leadingeikgen}
S_l(E,b) \approx 1+ \sum_{h=1}^{\infty} \frac{\mathcal{A}^{(1)}_{h}(E,b)}{2E} = e^{i \chi^{(1)}(E,b)}  \text{ ,}
\end{equation}
where $\mathcal{A}^{(1)}_{h}(E,b)$ is the leading energy contribution of the amplitude with $h$ boundaries and $\chi^{(1)}(E,b) = \mathcal{A}^{(1)}_{h=1}(E,b)/(2 E)$ is called the leading eikonal. Note that $S_l(E,b)$ captures all the information in the leading energy term of all amplitudes. We can write something similar for the subleading energy contribution by starting from $h=2$ and summing all the subleading contributions of the amplitudes at each number of boundaries. In this case we have the subleading eikonal given by $\chi^{(2)}(E,b) = \mathcal{A}^{(2)}_{h=2}(E,b)/(2 E)$ where $\mathcal{A}^{(2)}_{h}(E,b)$ is the subleading contribution to the amplitude with $h$ boundaries.

In the following we want to generalise the construction of the S-matrix in the eikonal approximation to include more general situations as for instance the presence of inelastic processes. Traditionally, for elastic processes, we write, including all contributions to all orders,
\begin{equation}
S(E,b) \approx \text{exp}\left({i \chi^{(1)}(E,b) + i \chi^{(2)}(E,b) + \ldots}\right) \text{ ,}
\label{Sela}
\end{equation}
where $\chi^{(1)}(E,b)$ is the leading eikonal and $\chi^{(2)}(E,b)$ is the subleading eikonal mentioned above. In subsection \ref{elasticeik} we show that this statement holds in the case of elastic dilaton scattering that we have already studied. In order to study this let us  write the tree-level ($h=1$) and one-loop ($h=2$) amplitudes as,
\begin{eqnarray}
\frac{i \mathcal{A}_{1}(E,b)}{2E} &=& i(N T_p \kappa_D) (A_{h=1}^{(1)}(b) E + A_{h=1}^{(2)}(b) E^{0} + \ldots) \label{diskampinE} \\
\frac{i \mathcal{A}_{2}(E,b)}{2E} &=& i(N T_p \kappa_D)^2 (A_{h=2}^{(1)}(b) E^2 + A_{h=2}^{(2)}(b) E + A_{h=2}^{(3)}(b) E^{0} + \ldots) \label{annampinE} \text{ ,}
\end{eqnarray}
where we have divided by $\frac{1}{\sqrt{2E}}$ for each of the two external particles involved and where the $A$ symbols correspond to $\mathcal{A}/2E$ where the dependence on energy has been factored out\footnote{We also note here that we are using the amplitudes once stripped of factors of $i (2\pi)^{p+1} \delta^{p+1}(k_1+k_2)$.}. Note here that in order to express the leading contributions as an exponential of the leading eikonal we have $i A_{h=2}^{(1)}(b)=-\frac{1}{2}(A_{h=1}^{(1)}(b))^2$. 

In equations (\ref{diskampinE}) and (\ref{annampinE}) we have also allowed for terms of order $E^0$ that, as we will see, are not present in the elastic dilaton scattering, but appear in the inelastic dilaton to RR scattering. We would like to extend the construction of the S-matrix in the eikonal approximation when these extra terms are present. Our proposal is that, in this more general case, (\ref{Sela}) is written as follows,
\begin{eqnarray}
S(E,b)  &=& \text{exp}\left[{\frac{i}{2} (\chi^{(1)}(E,b) +  \chi^{(2)}(E,b) + \ldots)}\right] \left(1 + T(E,b) \right) \times \nonumber \\
&& \;\;\; \text{exp}\left[{\frac{i}{2} (\chi^{(1)}(E,b) +  \chi^{(2)}(E,b) + \ldots)}\right] \text{ ,} \label{fullInelEik} 
\end{eqnarray}
where $\chi^{(1)}(E,b)$ and $\chi^{(2)}(E,b)$ are the leading eikonal and subleading eikonal respectively. The symbol $T(E,b)$ corresponds to all the non-diverging contributions to the amplitudes with any number of boundaries. For example the first contribution to $T(E,b)$ is $A_{h=1}^{(2)}(b)$; the first contribution to the tree-level dilaton to dilaton scattering process that does not grow with $E$. We have written \eqref{fullInelEik} in this way to account for when the eikonal and subleading eikonal behave as operators instead of phases. As can be seen from \cite{D'Appollonio:2013hja}, in string theory eikonal operators become important and it could therefore be useful for future considerations to be aware of this fact. In the cases considered in this paper the eikonal operators behave as phases and one can therefore recombine the exponentials. From the definitions above we see that to properly define the subleading eikonal we need,

\begin{equation}
i \frac{\chi^{(2)}(E,b)}{(NT_p \kappa_D)^2} = i A_{h=2}^{(2)}(b) E - \left( \frac{1}{2}i A_{h=1}^{(2)}(b) i A_{h=1}^{(1)}(b) E + \frac{1}{2}i A_{h=1}^{(1)}(b) i A_{h=1}^{(2)}(b) E \right) \;, \label{subleadingEik}
\end{equation}
where all the symbols have been defined above and we note that $A_{h=2}^{(2)}(b)$ represents the full subleading energy contribution derived from the one-loop amplitude in \eqref{annampinE}. Note that we have written (\ref{subleadingEik}) in the most general way possible accounting for the possibility that $i A_{h=1}^{(1)}(b)$ is an operator instead of a phase. In the cases we consider in this paper the eikonal operators become phases and the equation above reads $i \chi^{(2)}(E,b) = (NT_p \kappa_D)^2 \left(i A_{h=2}^{(2)}(b) E - i A_{h=1}^{(2)}(b) i A_{h=1}^{(1)}(b) E\right) $. \\

\subsection{Elastic Contributions to the Eikonal} \label{elasticeik}

We will calculate and discuss some explicit results in the high-energy limit for the interactions discussed in section \ref{2braneamps} for elastic dilaton scattering and show how this relates to the elastic eikonal scattering amplitude framework discussed at the start of this section.

\subsubsection{RR Sources} \label{AttachRRhighE}

The first and second terms of \eqref{FullRRSTAmp2} do not contribute to the high-energy limit. The first term is trivially $E^0$ as can be seen from the explicit expression for $\mathcal{I}_2$ in appendix \ref{integralref2prop}. The second term is more subtle but is also not of $\mathcal{O}(E^2)$ due to the extra propagator present in the integrals (the $1/u$ and $1/t$) which brings down a factor of $1/E$ after performing the integral, $\mathcal{I}_3$. The remaining terms we have are,

\begin{eqnarray}
i \mathcal{A}^{\rm dd \; (2)}_{h=2} &\approx& - i(N T_p \kappa_D)^2 \left( \frac{4}{s} k_{1 \mu} k_{2 \nu} \mathcal{I}_{2}^{\mu \nu} + 2 E^2 \mathcal{I}_{2} \right) \text{ .} 
\end{eqnarray}
Substituting the results for the various integrals,
\begin{equation}
i \mathcal{A}^{\rm dd \; (2)}_{h=2} \approx i(N T_p \kappa_D)^2 E^2 \frac{1}{(4 \pi)^{\frac{D-p-1}{2}}} \frac{\Gamma{\left( \frac{3-D+p}{2} \right)} \Gamma^2{\left( \frac{D-p-1}{2} \right)}}{\Gamma{\left( D-p-1 \right)}}  (q^2_{\perp})^{\frac{D-p-5}{2}} (4(D-p-2)-2) \text{ .} \label{finalampST_RR}
\end{equation}
Note that here and throughout this and the following section the $\approx$ signifies that we have dropped some terms that are subleading in energy which arise from performing the integrals. 

\subsubsection{Graviton Sources} \label{AttachGravhighE}

We can now substitute the relevant results for the integrals in \eqref{FullGGSTAmp} and identify which terms contribute to each power of energy. We have for the leading contribution,

\begin{eqnarray}
i \mathcal{A}^{\rm dd \; (1)}_{h=2} &=& i (N T_p \kappa_D)^2 4 E^4 \mathcal{I}_3^{(1)} \nonumber \\
&\approx& - (N T_p \kappa_D)^2 E^3 \frac{2 \sqrt{\pi}}{(4\pi)^{\frac{D-p-1}{2}}} \frac{\Gamma{\left(\frac{6-D+p}{2} \right)}  \Gamma^2\left(\frac{D-p-4}{2}\right)}{\Gamma(D-p-4)}(q^2_{\perp})^{\frac{D-p-6}{2}}  \text{ .} \label{GGAmpHighEu}
\end{eqnarray}
Note that in the last line we have used the solution for the leading energy contribution of $\mathcal{I}_3$ in appendix \ref{integralref} which we have denoted as $\mathcal{I}_3^{(1)}$. This is the only contribution at leading order in energy. The u- and t-channel diagrams which produce this leading contribution also have subleading contributions arising from the subleading term in $\mathcal{I}_3$,
\begin{eqnarray}
i \mathcal{A}^{\rm dd \; (2)}_{h=2} &=& i (N T_p \kappa_D)^2 4 E^4 \mathcal{I}_3^{(2)} \nonumber \\
&=& - i (N T_p \kappa_D)^2 E^2 \frac{2}{(4\pi)^{\frac{D-p-1}{2}}}  \frac{\Gamma{\left(\frac{5-D+p}{2} \right)} \Gamma^2(\frac{D-p-3}{2})}{\Gamma(D-p-4)} (q^2_{\perp})^{\frac{D-p-5}{2}} \;,
\label{GGAmpSubEu}
\end{eqnarray} 
where $\mathcal{I}_3^{(2)}$ the subleading energy contribution to $\mathcal{I}_3$. The other subleading contributions that arise from the second and third terms in \eqref{FullGGSTAmp} are,
\begin{eqnarray}
i \mathcal{A}^{\rm dd \; (2)}_{h=2} &=& i (N T_p \kappa_D)^2 \left(- \frac{2(D-2p-4)}{D-2} E^2 \mathcal{I}_2 + \frac{2(D-p-3)(1+p)}{D-2} \frac{1}{s} k_{1 \mu} k_{2 \nu}\mathcal{I}_{2}^{\mu \nu} \right) \nonumber \\
&\approx & i (N T_p \kappa_D)^2 E^2 \left(\frac{4(D-2p-4)(D-p-2)}{D-2} + \frac{(D-p-3)(1+p)}{D-2} \right) \nonumber \\
&& \times \frac{1}{(4\pi)^{\frac{D-p-1}{2}}}  \frac{\Gamma{\left( \frac{3-D+p}{2} \right)} \Gamma^2{\left( \frac{D-p-1}{2} \right)}}{ \Gamma{\left( D-p-1 \right)}} (q^2_{\perp})^{\frac{D-p-5}{2}} \text{ ,} \label{GGAmpHighEs}
\end{eqnarray}
and,
\begin{eqnarray}
i \mathcal{A}^{\rm dd \; (2)}_{h=2} &=& i (N T_p \kappa_D)^2 E^2 \frac{4 (D-p-3)}{D-2} \mathcal{I}_2 \nonumber \\
&=& - i (N T_p \kappa_D)^2 E^2 \frac{8(D-p-3)(D-p-2)}{D-2} \frac{1}{(4\pi)^{\frac{D-p-1}{2}}} \nonumber \\
&& \times  \frac{\Gamma{\left( \frac{3-D+p}{2} \right)} \Gamma^2{\left( \frac{D-p-1}{2} \right)}}{ \Gamma{\left( D-p-1 \right)}} (q^2_{\perp})^{\frac{D-p-5}{2}} \text{ ,} \label{GGAmpHighEc}
\end{eqnarray}
where we have separated the second and third terms of \eqref{FullGGSTAmp} into \eqref{GGAmpHighEs} and \eqref{GGAmpHighEc} purposefully in order to be able to more easily compare with the results obtained in section  \ref{lntoeikonal}.

\subsubsection{Dilaton Sources}\label{AttachDilhighE}

The only term contributing to the leading energy behaviour in this case is,
\begin{eqnarray}
i\mathcal{A}^{\rm dd \; (2)}_{h=2} &=& - 2 i \kappa_D^2 \left( N T_p \frac{a(D)}{\sqrt{2}} \right)^2  \frac{1}{s} k_{1 \mu} k_{2 \nu}\mathcal{I}_{2}^{\mu \nu} \nonumber \\
&\approx & i (N T_p \kappa_D)^2 \left(\frac{a(D)}{\sqrt{2}} \right)^2 E^2 \frac{1}{(4\pi)^{\frac{D-p-1}{2}}}  \frac{\Gamma{\left( \frac{3-D+p}{2} \right)} \Gamma^2{\left( \frac{D-p-1}{2} \right)}}{ \Gamma{\left( D-p-1 \right)}} (q^2_{\perp})^{\frac{D-p-5}{2}} \label{finalampFT_Dil}
\end{eqnarray}
where we have used the kinematics outlined in section \ref{kinematics}.

\subsubsection{Eikonal Scattering} \label{ElasticEikDiscussion}

We can use the results derived above to explicitly show that \eqref{leadingeikgen} holds for the elastic scattering of dilatons from D-branes. Writing the leading energy behaviour of the tree-level and one-loop amplitudes in the form of \eqref{diskampinE} and \eqref{annampinE} respectively and by converting these expressions into impact parameter space using \eqref{impoformu}, we find for the tree-level amplitude,
\begin{eqnarray}
i A_{h=1}^{(1)\,e}(b) &=& \frac{i}{4 \pi^{\frac{D-p-2}{2}}} \frac{\Gamma \left(\frac{D-p-4}{2} \right) }{b^{D-p-4}}\;, \label{elEips} \\
i A_{h=1}^{(2)\,e}(b) &=& 0 \;,
\end{eqnarray}
and for the one-loop amplitude,
\begin{eqnarray}
i A_{h=2}^{(1)\,e}(b) &=& - \frac{1}{32 \pi^{D-p-2}} \frac{\Gamma^2 \left(\frac{D-p-4}{2} \right) }{b^{2D-2p-8}}\;, \\
i A_{h=2}^{(2)\,e}(b) &=& i \frac{1}{16 \pi^{D-p-3/2}} \frac{\Gamma^2 \left(\frac{D-p-3}{2} \right) \Gamma \left(\frac{2D-2p-7}{2} \right)}{\Gamma \left(D-p-4 \right)} \frac{1}{b^{2D-2p-7}} \;,
\end{eqnarray}
where here we are focusing on the elastic component as reminded by the superscript $e$. We note that for the one-loop amplitude the contributions to $A_{h=2}^{(2)\,e}(b)$ arising from \eqref{finalampST_RR}, \eqref{GGAmpHighEs}, \eqref{GGAmpHighEc} and \eqref{finalampFT_Dil} sum to zero. This means that the only contribution to the subleading eikonal arises from the subleading contribution to the leading energy contribution, \eqref{GGAmpSubEu}, where we recall that $\mathcal{I}_3$ has contributions at different powers of $E$.

We can now easily confirm that $i A_{h=2}^{(1)\,e}(b)=-\frac{1}{2}(A_{h=1}^{(1)\,e}(b))^2$ as required in order to see the exponentiation of the leading eikonal $\chi^{(1)}(E,b)$ in the elastic channel. We therefore find that the elastic dilaton scattering process we have considered behaves as predicted by the leading eikonal expression \eqref{leadingeikgen}. 

\subsection{Inelastic Contributions to the Eikonal} \label{inelasticeik}

As we've done in section \ref{elasticeik} for the elastic dilaton scattering process, we can find the leading energy behaviour of the inelastic scattering of a dilaton and an RR field from the stack of D-branes that we studied in section \ref{ineldilRRallE}. Looking at the leading energy contribution of \eqref{eq:dRgRf} we find,
\begin{eqnarray}
&& i \mathcal{A}_{h=2}^{\rm dR \; (2)} =  i (N T_p \kappa_D)^2 8 a(D)( q \cdot C^{(2)})^{1 \ldots p} E^3 \mathcal{I}_3^{(1)} \nonumber \\
&& \approx   - (N T_p \kappa_D)^2 2 a(D) E^2 ( q \cdot C^{(2)})^{1 \ldots p} \frac{2 \sqrt{\pi}}{(4\pi)^{\frac{D-p-1}{2}}} \frac{\Gamma{\left(\frac{6-D+p}{2} \right)}  \Gamma^2\left(\frac{D-p-4}{2}\right)}{\Gamma(D-p-4)}(q^2_{\perp})^{\frac{D-p-6}{2}} \;
\label{dRRAmpHighE}
\end{eqnarray}
We can now apply the prescription outlined in \eqref{fullInelEik} to \eqref{dRRAmpHighE}. Writing the tree-level amplitude \eqref{eq:dil-RR0} in the form of \eqref{diskampinE} we find that
\begin{eqnarray}
i A_{h=1}^{{\rm dR} (1)}(b) &=& 0 \\
T(E,b) \approx i A_{h=1}^{{\rm dR} (2)}(b) &=& i \frac{a(D)}{4} ( q \cdot C)^{1 \ldots p} \frac{1}{\pi^{\frac{D-p-2}{2}}} \frac{\Gamma \left(\frac{D-p-4}{2} \right) }{b^{D-p-4}} \;. \label{inelE0ips}
\end{eqnarray}
The other ingredients we need are $A_{h=2}^{(1)}(b)$ and $A_{h=2}^{(2)}(b)$ that we can read by comparing~\eqref{dRRAmpHighE} and \eqref{annampinE},
\begin{eqnarray}
i A_{h=2}^{{\rm dR} (1)}(b) &=& 0 \\
i A_{h=2}^{{\rm dR} (2)}(b) &=& - \frac{a(D)}{16} ( q \cdot C)^{1 \ldots p} \frac{1}{\pi^{D-p-2}} \frac{\Gamma^2 \left(\frac{D-p-4}{2} \right) }{b^{2D-2p-8}} \;.
\end{eqnarray}
We then need to calculate $i A_{h=1}^{(1) \,e}(b) i A_{h=1}^{{\rm dR} (2)}(b)$ as this will show us what to subtract in order to obtain the well defined subleading eikonal $\chi^{(2)}(E,b)$, including the inelastic contributions discussed above. We note here that although this inelastic process does not contribute to the total $A_{h=1}^{(1)}(b)$ we have to take into account the contribution from the elastic processes described in section \ref{ElasticEikDiscussion}.  We can easily verify by using \eqref{elEips} and \eqref{inelE0ips} that,
\begin{equation}
  \label{eq:dRchi0}
  i A_{h=2}^{{\rm dR} (2)}(b)- i A_{h=1}^{(1) \,e}(b) i A_{h=1}^{{\rm dR} (2)}(b) = 0
\end{equation}
and so we see that the inelastic dilaton to RR channel does not contribute to the subleading eikonal~\eqref{subleadingEik}, as the corresponfing component of $i A_{h=2}^{(2)}(b) E - i A_{h=1}^{(1)}(b) i A_{h=1}^{(2)}(b) E$ vanishes.


\section{Alternative Computation of the Leading and Subleading Eikonal}
\label{lntoeikonal}

In this section we discuss a more conventional way to compute the elastic scattering of a dilaton from a stack of D$p$-branes both in Einstein gravity and in a theory of gravity extended to include the dilaton and RR fields.  We will take the high energy limit and extract the leading and subleading eikonal, which agrees with the ones computed in the previous section. The leading eikonal is obtained from the tree diagram corresponding to the exchange of a graviton, while the subleading eikonal is derived from a number of one-loop  diagrams that depend on which theory of gravity we consider. Since three of the one-loop diagrams are most easily obtained by first computing the one-point graviton amplitude and then attaching the three-point vertex containing two dilatons and one graviton, in the first subsection we compute the one-point amplitudes for the graviton, dilaton and RR field at the tree and one-loop level and we show that they are  directly related to the large distance behaviour of the classical solution describing the 
D$p$-branes to which the graviton, dilaton and RR field are coupled. In the second subsection we compute the contribution of the various field theory diagrams to the elastic dilaton scattering and from them we extract the leading and subleading eikonal. 

\subsection{One-point Amplitudes and the Classical Solution}
\label{1point}

In this subsection we will write the one-point functions for the graviton, dilaton and RR field in the gravity theory described by the bulk action given in \eqref{eq:bulkaction} and the boundary action given by \eqref{branecoup} as in the previous sections. Using these two actions one can compute the contribution to the one-point amplitude of the diagram with the 3-graviton vertex yielding\footnote{In all one-point amplitudes we omit to explicitly write a $\delta$-function that constrains the longitudinal component of the momentum to be vanishing.},

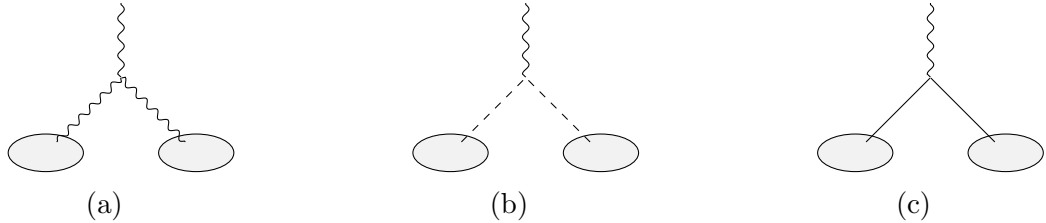
\begin{figure}[h]
  \begin{subfigure}[t]{0.3\textwidth}
    \centering
    \begin{tikzpicture}
		\begin{feynman}
			\vertex (a) at (-1,-2) {};
			\vertex (b) at ( 1,-2) {};
			\vertex (c) at (-2.5, 0) {};
			\vertex (d) at ( 2.5, 0) {};
			\vertex[circle,inner sep=0pt,minimum size=0pt] (e) at (0, -1) {};
			\vertex[circle,inner sep=0pt,minimum size=0pt] (m) at (0, 0) {};
			\draw[fill=light-gray] (-1,-2) ellipse (0.5cm and 0.25cm);
			\draw[fill=light-gray] (1,-2) ellipse (0.5cm and 0.25cm);       
			\diagram* {
			(m) -- [boson] (e),
			(a) -- [boson] (e),
			(b) -- [boson] (e),
			};
		\end{feynman}
    \end{tikzpicture}
    \caption{}
    \label{fig:5a}
  \end{subfigure}
  \quad
  \begin{subfigure}[t]{0.3\textwidth}
    \centering
    \begin{tikzpicture}
		\begin{feynman}
			\vertex (a) at (-1,-2) {};
			\vertex (b) at ( 1,-2) {};
			\vertex (c) at (-2.5, 0) {};
			\vertex (d) at ( 2.5, 0) {};
			\vertex[circle,inner sep=0pt,minimum size=0pt] (e) at (0, -1) {};
			\vertex[circle,inner sep=0pt,minimum size=0pt] (m) at (0, 0) {};
			\draw[fill=light-gray] (-1,-2) ellipse (0.5cm and 0.25cm);
			\draw[fill=light-gray] (1,-2) ellipse (0.5cm and 0.25cm);       
			\diagram* {
			(m) -- [boson] (e),
			(a) -- [scalar] (e),
			(b) -- [scalar] (e),
			};
		\end{feynman}
    \end{tikzpicture}
    \caption{}
    \label{fig:5b}
  \end{subfigure}
  \quad
  \begin{subfigure}[t]{0.3\textwidth}
    \centering
    \begin{tikzpicture}
		\begin{feynman}
			\vertex (a) at (-1,-2) {};
			\vertex (b) at ( 1,-2) {};
			\vertex (c) at (-2.5, 0) {};
			\vertex (d) at ( 2.5, 0) {};
			\vertex[circle,inner sep=0pt,minimum size=0pt] (e) at (0, -1) {};
			\vertex[circle,inner sep=0pt,minimum size=0pt] (m) at (0, 0) {};
			\draw[fill=light-gray] (-1,-2) ellipse (0.5cm and 0.25cm);
			\draw[fill=light-gray] (1,-2) ellipse (0.5cm and 0.25cm);       
			\diagram* {
			(m) -- [boson] (e),
			(a) -- (e),
			(b) -- (e),
			};
		\end{feynman}
    \end{tikzpicture}
    \caption{}
    \label{fig:5c}
  \end{subfigure}
  \caption{The various contributions to the one-point function at subleading order used to construct the classical solution. Figure \ref{fig:5a} is the contribution with the 3-graviton vertex and figures \ref{fig:5b} and \ref{fig:5c} are the contributions with RR fields and dilaton sources respectively. As before the solid lines correspond to dilatons, the wavy lines correspond to gravitons and the dashed lines correspond to RR fields.}
  \label{fig:5}
\end{figure}

\begin{eqnarray}
&&  \langle h_{\lambda \tau} \rangle_{(\ref{fig:5a})} = N^2 \kappa_D
J_h^2 
\left[\frac{ |q_{\perp}|^{D-p-5}  }{
(4\pi)^{ \frac{D-p-1}{2} }  } \Gamma \left(\frac{3-D+p}{2}\right) 
\frac{\Gamma^2 (\frac{D-p-1}{2})}{
\Gamma (D-p-1)} \right] \nonumber \\
&& \times \Big\{ \frac{(D-3-p) (p+1)}{2 (D-2)} 
\Big[ \eta_{\perp \lambda \tau} - (3-D+p) \frac{q_{\perp \lambda} 
q_{\perp \tau}}{q_{\perp}^2}   \Big] \nonumber \\
&& - 2 \frac{(D-p-2) (p+1) (D-p-3)}{D-2} 
\frac{q_{\perp \lambda} 
q_{\perp \tau}}{q_{\perp}^2}   \nonumber \\
&& - \frac{2(D-p-2)(D-p-3)^2}{(D-2)^2} \eta_{\parallel \lambda \tau}
- \frac{2 (D-p-2) (p+1)^2}{(D-2)^2} \eta_{\perp \lambda \tau}
\Big\} \;,
\label{3gra1}
\end{eqnarray}
The sum of the contributions from the diagrams with the dilaton and the RR field is given by
\begin{eqnarray}
&&\langle h \rangle_{(\ref{fig:5b}) + (\ref{fig:5c})} = N^2 \kappa_D 
\left[\frac{ |q_{\perp}|^{D-p-5}  }{
(4\pi)^{ \frac{D-p-1}{2} }  } \Gamma \left(\frac{3-D+p}{2}\right) 
\frac{\Gamma^2 (\frac{D-p-1}{2})}{
\Gamma (D-p-1)} \right]  \nonumber \\
&& \times \Big\{ \frac{J_{\phi}^2}{2}
\Big[ \eta_{\perp \lambda \tau} - (3-D+p) \frac{q_{\perp \lambda} 
q_{\perp \tau}}{q_{\perp}^2}   \Big]  \label{3gra2} \\
&& + \frac{\mu_p^2}{2} \Big[  - \frac{2(D-p-2) (D-p-3)}{D-2} 
 \eta_{\parallel \lambda \tau} + \frac{2 (D-p-2) (p+1)}{D-2} 
\eta_{\perp \lambda \tau}  \nonumber \\ \nonumber
&& - \Big[ \eta_{\perp \lambda \tau} - (3-D+p) \frac{q_{\perp \lambda} 
q_{\perp \tau}}{q_{\perp}^2}   \Big]  \Big]  \Big\} \;.
\end{eqnarray}
Summing the  three contributions  we get
\begin{eqnarray}
&&\langle h \rangle_{(\ref{fig:5a}) + (\ref{fig:5b}) + (\ref{fig:5c})} =N^2 \kappa_D 
\left[\frac{ |q_{\perp}|^{D-p-5}  }{
(4\pi)^{ \frac{D-p-1}{2} }  } \Gamma \left(\frac{3-D+p}{2}\right) 
\frac{\Gamma^2 (\frac{D-p-1}{2})}{
\Gamma (D-p-1)} \right] \nonumber \\   
&& \times \Big\{  - 2 \frac{(D-p-2) (p+1) (D-p-3)}{D-2} 
\frac{q_{\perp \lambda} 
q_{\perp \tau}}{q_{\perp}^2}   \nonumber \\
&& - \frac{2(D-p-2)}{D-2}  (D-p-3)\eta_{\parallel \lambda \tau} 
\Big[ J_h^2 \frac{D-p-3}{D-2} + \frac{\mu_p^2}{2} \Big] 
\nonumber \\
&& - \frac{2 (D-p-2)}{D-2} (p+1) \eta_{\perp \lambda \tau} \Big[ J_h^2 
\frac{p+1}{D-2} - 
\frac{\mu_p^2}{2} \Big] \Big\} \;,
\label{htotal}
\end{eqnarray}
where from the expressions for $J_{\phi}$, $\mu_p$ and $a(D)$ defined in appendix \ref{appFeynRules}, we have used the fact that the following quantity vanishes,
\begin{eqnarray}
 \frac{(D-p-3)(p+1)}{2(D-2)}J_h^2 + 
\frac{J_{\phi}^2}{2}- \frac{\mu_p^2}{2}  = 0 \;.
\label{gra5r}
\end{eqnarray}
We neglect for a moment the term in the second line of \eqref{htotal} that corresponds to a gauge transformation  of the metric as we will discuss it in subsection \ref{Einstein} where we will see that it must be neglected if we want the metric in the harmonic gauge.

Going from momentum to impact parameter space, (\ref{htotal}) becomes
\begin{eqnarray}
&&\langle \tilde{h}_{\mu \nu} \rangle_{(\ref{fig:5a}) + (\ref{fig:5b}) + (\ref{fig:5c})} = \frac{N^2 \kappa_D}{D-2} 
\left( \frac{1}{(D-p-3) \Omega_{D-p-2} r^{D-p-3}}
  \right)^2 \nonumber \\
&& \times
 \left\{ \eta_{\parallel \mu \nu} (D-p-3) \left[  J_h^2 \frac{D-p-3}{D-2}
 +\frac{ \mu_p^2}{2} \right] + (p+1) \eta_{\perp \mu \nu}
 \left[  J_h^2 \frac{p+1}{D-2}
 -\frac{ \mu_p^2}{2} \right]  \right\} \;, \nonumber \\
 \label{gra7r}
 \end{eqnarray}
 where we note that the tilde signifies the Fourier transform to impact parameter space. The expression in impact parameter space can be obtained by using \eqref{impoformu}. Inserting the explicit quantities (\ref{gra7r}) becomes
 \begin{eqnarray}
 &&\langle 2 \kappa_D \tilde{h}_{\mu \nu} \rangle_{(\ref{fig:5a}) + (\ref{fig:5b}) + (\ref{fig:5c})} = 
\frac{1}{2}\left(\frac{R_p}{r} \right)^{2(D-p-3)} 
\nonumber \\
&& \times
 \left\{ \eta_{\parallel \mu \nu} \frac{D-p-3}{D-2} \left[   \frac{D-p-3}{D-2}
 + 1 \right] + \frac{p+1}{D-2}  \eta_{\perp \mu \nu}
 \left[   \frac{p+1}{D-2}  - 1 \right]  \right\} \;, \nonumber \\
\label{gra7b}
\end{eqnarray}
where we have introduced the following quantity,
\begin{eqnarray}
 \frac{2 N \kappa_D T_p}{(D-p-3) \Omega_{D-p-2} r^{D-p-3}}
\equiv \left(\frac{R_p}{r} \right)^{D-p-3}  ~~;~~
 \Omega_d \equiv \frac{ 2 \pi^{\frac{d+1}{2} }}{\Gamma (
\frac{d+1}{2})} \;.
\label{Rp}
\end{eqnarray}
We note that (\ref{gra7b}) provides the  total one-loop contribution to the one-point graviton amplitude. The tree contribution can
also be easily computed from the bulk and boundary actions yielding,
\begin{eqnarray}
\langle 2 \kappa_D \tilde{h}_{\mu \nu} (x) \rangle _{1}= - 
\left(\frac{R_p}{r} \right)^{D-p-3} 
 \left( \frac{D-p-3}{D-2} \eta_{\mu \nu}^{\parallel} 
- \frac{p+1}{D-2} \eta_{\mu \nu}^{\perp} \right) \;,
\label{gra13}
\end{eqnarray}
which is the Fourier transform of the following amplitude in momentum space,
\begin{eqnarray}
\langle h_{\mu \nu} \rangle_{1} = - \frac{N T_p}{q_\perp^2} \left( \frac{D-p-3}{D-2} \eta_{\parallel \mu \nu} - \frac{p+1}{D-2} \eta_{\perp\mu \nu} \right) \;.
\label{gra13f}
\end{eqnarray}
Note that we are using the same notation as in section \ref{2braneamps} with subscripts $1$ and $2$ representing tree diagrams and one-loop diagrams respectively. In an extended gravity theory also containing the dilaton and the RR field we have to include the one-point amplitude for the dilaton and the RR field. The one-loop one-point amplitude for the dilaton is given  by the sum of two diagrams. One with the vertex containing two dilatons and one graviton and the other with the vertex with one dilaton and two RR fields.
It turns out that the first diagram is vanishing while the second one gives,
\begin{eqnarray}
\langle \phi \rangle_{2} = 
\frac{a(D)  \sqrt{2} N^2 \kappa_D   }{2 }  \mu_p^2   
(2-D+p)
\left[\frac{ |q_{\perp}|^{D-p-5}  }{
(4\pi)^{ \frac{D-p-1}{2} }  } \Gamma \left(\frac{3-D+p}{2}\right) 
\frac{\Gamma^2 (\frac{D-p-1}{2})}{
\Gamma (D-p-1)} \right] \;,
\label{OD15}
\end{eqnarray} 
where the dilaton field has been canonically normalised. From (\ref{OD15}) we can go to impact parameter space,
\begin{eqnarray}
&&\langle \sqrt{2} \kappa_D \tilde{\phi} \rangle_{2} = 
\frac{ a(D) }{4 }
 \left( \frac{2 N \kappa_D T_p}{(D-p-3) \Omega_{D-p-2} r^{D-p-3}} 
\right)^2 = \frac{ a(D) }{4 }  \left(\frac{R_p}{r} \right)^{2(D-p-3)} \;,
\label{OD18}
\end{eqnarray}
where we have used that $\mu_p = \sqrt{2} T_p$.  We also have the tree diagram that in momentum space gives the following contribution
\begin{eqnarray}
 \langle \phi \rangle_{1}  = N
J_\phi \frac{1}{q_\perp^2} \;,
\label{OD23}
\end{eqnarray} 
which in impact parameter space becomes,
\begin{eqnarray}
\langle  \sqrt{2} \kappa_D \tilde{\phi} \rangle_{1} = \frac{J_\phi}{\sqrt{2}T_p} 
\left(\frac{R_p}{r} \right)^{D-p-3} 
=- \frac{a(D)}{2}   \left(\frac{R_p}{r} \right)^{D-p-3} \;.
\label{OD24}
\end{eqnarray} 
The one-loop one-point amplitude for the RR fields gets a contribution from two diagrams; one with the vertex involving two RR fields and one graviton and the other involving again two RR fields and a dilaton. The sum of the two is equal, in momentum space, to
\begin{eqnarray}
\langle C_{01 \dots p}\rangle_{2} =  4 N^2 T_p \kappa_D \mu_p (D-p-2)  \left[\frac{ |q_{\perp}|^{D-p-5}  }{
(4\pi)^{ \frac{D-p-1}{2} }  } \Gamma \left(\frac{3-D+p}{2}\right) 
\frac{\Gamma^2 (\frac{D-p-1}{2})}{
\Gamma (D-p-1)} \right] \;,
\label{RR15}
\end{eqnarray} 
which in impact parameter space becomes,
\begin{eqnarray}
&&\langle \tilde{C}_{01 \dots p}\rangle_{2} = - 4 N^2 T_p \kappa_D \mu_p  \frac{1}{2}   
  \left( \frac{1}{(D-p-3) \Omega_{D-p-2} r^{D-p-3}} \right)^2 \;,
\label{RR16}
\end{eqnarray} 
where the field $C_{01 \dots p}$ is canonically normalised. In order to compare this with the classical solution, we need the quantity,
\begin{eqnarray}
&&\langle \sqrt{2} \kappa_D \tilde{C}_{01 \dots p}\rangle_{2}  = -  
  \left( \frac{2 N \kappa_D T_p}{(D-p-3) \Omega_{D-p-2} r^{D-p-3}} \right)^2  =- \left(\frac{R_p}{r} \right)^{2(D-p-3)}  \;.
\label{RR18}
\end{eqnarray} 
The tree diagram can also be easily computed, we find
\begin{eqnarray}
\langle \sqrt{2} \kappa_D \tilde{C}_{01 \dots p} \rangle_{1}= 
-\frac{2 N T_p \kappa_D}{(D-p-3) \Omega_{D-p-2} r^{D-p-3}} = \left(\frac{R_p}{r} \right)^{D-p-3} \;.
\label{RR22}
\end{eqnarray}

The previous diagrammatic results, obtained for the various one-point amplitudes, can be compared with the large distance expansion of the classical solution. It turns out that  the tree diagrams reproduce the first correction to the flat limit  of the classical solution when $r \rightarrow \infty$, while the one-loop diagrams reproduce the subleading correction to the flat limit. The classical solution is given by \cite{Duff:1994an},
\begin{eqnarray}
&& ds^2 \equiv g_{\mu \nu} dx^\mu d x^\nu =  [H(r)]^{-\frac{D-p-3}{D-2}  } dx_{\parallel}^2 + [H(r)]^{\frac{p+1}{D-2}}
 dx_{\perp}^2   \nonumber \\
 && {\rm e}^{- \sqrt{2} \kappa_D \phi} = ( H (r) )^{a (D)/2}~~;~~~\sqrt{2} \kappa_D C_{01 \dots p} = 1 -
  H^{-1} (r)  \;,
\label{ds2xy}
\end{eqnarray}
where
\begin{eqnarray}
H(r) = 1 + \left(\frac{R_p}{r} \right)^{D-p-3} \;.
\label{gra8}
\end{eqnarray}
Expanding the two terms appearing in the metric, we get
\begin{eqnarray}
&&[H(r)]^{-\frac{D-p-3}{D-2}  }  = 1 - \frac{D-p-3}{D-2} 
 \left(\frac{R_p}{r} \right)^{D-p-3} 
 \nonumber \\ 
 &&
 + \frac{1}{2} \frac{D-p-3}{D-2} \left( \frac{D-p-3}{D-2} +1 \right) \left(\frac{R_p}{r} \right)^{2(D-p-3)}   + \dots
\label{gra10x}
\end{eqnarray}
and
\begin{eqnarray}
&& [H(r)]^{\frac{p+1}{D-2}} = 1 + \frac{p+1}{D-2}  \left(\frac{R_p}{r} \right)^{D-p-3} 
 \nonumber \\ 
 && 
 + \frac{1}{2} \frac{p+1}{D-2} \left( \frac{p+1}{D-2} -1 \right)  \left(\frac{R_p}{r} \right)^{2(D-p-3)}  +\dots
\label{gra11x}
\end{eqnarray}
Remembering that in our notation $g_{\mu \nu} =\eta_{\mu \nu} + 2 \kappa_D h_{\mu \nu}$, we see that
for $r \rightarrow \infty$ we get the flat Minkowski metric. Then, comparing (\ref{ds2xy}), (\ref{gra10x}) and (\ref{gra11x}) with (\ref{gra13}) and (\ref{gra7b}), we see that the first correction to the flat space metric is given by the tree diagram of the one-point graviton amplitude, while the next correction is given by the one-loop diagram contribution to the one-point graviton amplitude. Expanding in a similar way the classical solution for the dilaton we get
\begin{eqnarray}
&& - \sqrt{2} \kappa_D \phi = \frac{a(D)}{2} \log \left(1+ \left(\frac{R_p}{r} \right)^{D-p-3} \right) 
\nonumber \\
&& = \frac{a(D)}{2} \left(  \left(\frac{R_p}{r} \right)^{D-p-3} - \frac{1}{2}  \left(\frac{R_p}{r} \right)^{2(D-p-3)}  +
\dots \right) \;.
\nonumber \\
\label{dilaFF}
\end{eqnarray}
These two terms are  reproduced by the tree diagram in (\ref{OD24}) and the one-loop term in (\ref{OD18}) respectively. Similarly expanding the solution for the RR field we get
\begin{eqnarray}
&& \sqrt{2} \kappa_D C_{01 \dots p} =  1 - H^{-1}  =   \left(\frac{R_p}{r} \right)^{D-p-3}  -
 \left(\frac{R_p}{r} \right)^{2(D-p-3)}  + \dots 
\label{dilRR}
\end{eqnarray}
Again we find that these two terms are equal to those in (\ref{RR22}) and (\ref{RR18}).

In conclusion, we have shown  that the various terms of the expansion of the classical solution
can be reproduced by computing the one-point amplitude of the corresponding fields.

\subsection{Elastic Dilaton Scattering in Extended Gravity}
\label{ElaExtGra}

In this subsection we compute the elastic dilaton scattering amplitude in an extended theory of gravity with a dilaton and an RR field as in section \ref{elasticeik}. It consists of one tree diagram and five one-loop diagrams. The tree diagram and the sum of three one-loop diagrams can be obtained directly from the one-point amplitude computed in the previous subsection by saturating it with the three-point amplitude of two dilatons and one graviton given in (\ref{ddgV}). For the tree diagram we find the following\footnote{In this case we also omit writing the factor $(2\pi)^{p+1} \delta^{(p+1)} (k_1 +k_2)$ of momentum conservation along the directions of the D$p$-brane.},
\begin{eqnarray}
i\mathcal{A}_{1}^{\rm dd} = i \frac{2 N T_p \kappa_D}{(-s)} \left(\frac{D-p-3}{D-2} (k_{1} \cdot k_2)_{\parallel}  - \frac{p+1}{D-2}  (k_{1} \cdot k_2)_{\perp}   \right)  =  i \frac{2 N T_p \kappa_D E^2}{(-s)} \;,
\label{EDS1}
\end{eqnarray}
where we have neglected terms without the pole at $s \sim 0$ as well as terms negligible at high energy (see kinematics in (\ref{k1k2})). We find that this is in agreement with \eqref{eq:dil-dil0}.

The first one-loop diagram corresponds to the separate exchange of two gravitons that  are then attached to the D$p$-branes. One gets
\begin{eqnarray}
i \mathcal{A}_{2, u}^{\rm ddgg} = i (N \kappa_D T_p)^2 4E^4 \int \frac{d^{D-p-1} k}{(2\pi)^{D-p-1}} \frac{1}{(k_1 - k)_{\perp}^2 k^2 (k_2 +k)^2_{\perp} } \;,
\label{EDS2}
\end{eqnarray}
where $k^2 \equiv -E^2 + k_{\perp}^2$. At high energy we obtain a leading term given by
\begin{eqnarray}
i \mathcal{A}_{2, u}^{\rm ddgg} \approx  -  \frac{2  (N \kappa_D T_p)^2E^3 \sqrt{\pi} }{  (4 \pi)^{\frac{D-p-1}{2}}} 
\frac{\Gamma{\left(\frac{6-D+p}{2} \right)} \Gamma^2(\frac{D-p-4}{2})}{\Gamma(D-p-4)}  (q^2_{\perp})^{\frac{D-p-6}{2}} \;,
\label{EDS3}
\end{eqnarray}
and a subleading term equal to
\begin{eqnarray}
i \mathcal{A}_{2, u}^{\rm ddgg} \approx - i \frac{(N \kappa_D T_p)^2  2E^2}{(4\pi)^{\frac{D-p-1}{2}}}
\frac{\Gamma (\frac{5-D+p}{2}) \Gamma^2 ( \frac{D-p-3}{2}) }{ 
\Gamma (D-p-4)} (q^2_{\perp})^{\frac{D-p-5}{2}} \;. 
\label{EDS4}
\end{eqnarray}
Comparing \eqref{EDS3} and \eqref{EDS4} with the equivalent results \eqref{GGAmpHighEu} and \eqref{GGAmpSubEu} derived in section \ref{HElimit} we again find agreement. The second diagram contains a vertex with two dilatons and two gravitons with the gravitons attached to the D-branes.  We find that
\begin{eqnarray}
&& i \mathcal{A}_{2, c}^{\rm ddgg}  = i   (N \kappa_D T_p)^2     
\int \frac{d^{D-p-1} k}{(2 \pi)^{D-p-1}} 
 \frac{1}{k^{2}_{\perp}  (q-k)^{2}_{\perp}}   \nonumber \\
&& \times \frac{D-3-p}{D-2} \left( -(p+1) (k_1 \cdot k_2)  + 
4 (k_{1} \cdot k_{2} )_{\parallel} \right) \;.
\label{EDS5}
\end{eqnarray}
In the high energy limit we can neglect the first term in the round bracket in the second line and we find,
\begin{eqnarray}
i \mathcal{A}_{2, c}^{\rm ddgg} \approx - i \frac{(N \kappa_D T_p)^2 8 E^2}{(4\pi)^{\frac{D-1-p}{2}}}  \frac{(D-p-3)(D-p-2)}{D-2}\,\,
\frac{ \Gamma( \frac{3-D+p}{2}) \Gamma^2( \frac{D-p-1}{2})}{
\Gamma ( D-p-3)} (q^2_{\perp})^{\frac{D-p-5}{2}}  \;.
\label{EDS6}
\end{eqnarray}
We can easily see that this is equivalent to \eqref{GGAmpHighEc}. Finally, the last three one-loop diagrams are obtained by saturating the one-point amplitudes in (\ref{3gra1}) and (\ref{3gra2}) with the vertex in (\ref{ddgV}).   Let us start with the one-loop diagram in (\ref{3gra1}). The term with $(k_1 \cdot k_2)$ in (\ref{ddgV})  and the second term in the second line and the term in the third line of (\ref{3gra1}) do not contribute at high energy. The remaining terms give,
\begin{eqnarray}
&&i \mathcal{A}_{2, s}^{\rm ddgg} \approx i \frac{(N \kappa_D T_p E)^2}{D-2} \frac{1}{(4\pi)^{\frac{D-p-1}{2}}}  \frac{ \Gamma (\frac{3-D+p}{2}) \Gamma^2 (\frac{D-p-1}{2} )}{ \Gamma (D-p-1)} (q^2_{\perp})^{\frac{D-p-5}{2}} \times
\nonumber \\
&& \left( (D-p-3)(p+1) + 4 (D-p-2) (D-2p -4)  \right) \;.
\label{EDS7}
\end{eqnarray}
Once again comparing this with our results from section \ref{HElimit} we see that the equation above is equivalent to \eqref{GGAmpHighEs}. Let us do the same analysis with (\ref{3gra2}). Again the term with $(k_1 \cdot k_2)$ in (\ref{ddgV}) does not contribute at high energy. Also the terms with $q_{\perp \lambda} q_{\perp \tau}$ do not contribute at high energy. We are therefore left with the following expression,
\begin{eqnarray}
&&i \mathcal{A}_{2, s}^{\rm ddRR} + i \mathcal{A}_{2, s}^{\rm dddd}  \approx i N^2 \kappa_D^2 E^2 \frac{1}{(4\pi)^{\frac{D-p-1}{2}}}  \frac{ \Gamma (\frac{3-D+p}{2}) \Gamma^2 (\frac{D-p-1}{2} )}{ \Gamma (D-p-1)} (q^2_{\perp})^{\frac{D-p-5}{2}} \times
\nonumber \\
&& \times  \left( J_\phi^2 + \mu_p^2 \left( 2(D-p-2) -1 \right) \right)  \;.
\label{EDS8}
\end{eqnarray}
Inserting the relevant expression for $J_\phi$ and using $\mu_p = \sqrt{2} T_p$,
\begin{eqnarray}
&& i \mathcal{A}_{2, s}^{\rm ddRR} + i \mathcal{A}_{2, s}^{\rm dddd}  \approx i (N \kappa_D T_p E)^2   \frac{1}{(4\pi)^{\frac{D-p-1}{2}}}  \frac{ \Gamma (\frac{3-D+p}{2}) \Gamma^2 (\frac{D-p-1}{2} )}{ \Gamma (D-p-1)} (q^2_{\perp})^{\frac{D-p-5}{2}} \times
\nonumber \\
&& \times \left[  \left(2 - \frac{(p+1)(D-p-3)}{D-2} \right) + \left( 4(D-p-2) -2 \right)  \right]  \;,
\label{EDS9}
\end{eqnarray}
where the first round bracket in the second line comes from the dilaton, while the second round bracket comes from the RR contribution. This can be compared to the sum of \eqref{finalampST_RR} and \eqref{finalampFT_Dil}, which agrees with what is written above.

The total eikonal, including both leading and subleading contributions, is defined as $\chi (b,E)=\chi^{(1)}(E,b)+\chi^{(2)}(E,b)$, where the $\chi^{(i)}(E,b)$ have been defined in section \ref{HElimit}. From the various expressions computed in this section we arrive at the following expression,
\begin{eqnarray}
&& \chi (b, E) = \frac{N \kappa_D^2 \tau_p}{4}  E\frac{\Gamma 
( \frac{D-p-4}{2})}{ \pi^{\frac{D-p-2}{2}} b^{D-p-4}} 
+\frac{(N \kappa_D^2 \tau_p)^2 \, E \,\Gamma^2 ( \frac{D-p-3}{2})
\Gamma (D-p - \frac{7}{2})}{16 \,\pi^{D-p - \frac{3}{2}} \Gamma (D-p-4) 
\,\,b^{2D-2p-7}} \nonumber \\
&& + \frac{(N \kappa_D^2 \tau_p)^2 \, E \, \Gamma (D-p - \frac{7}{2})
\Gamma^2 ( \frac{D-p-1}{2})}{16 (3+p-D) \Gamma (D-p-1)  
\,\,\pi^{D-p - \frac{3}{2}}\,\, b^{2D-2p-7}}   \nonumber \\
&& \times  
\Bigg\{ -  \frac{8(D-p-2) (D-p-3)}{D-2} \nonumber \\
&& +  \frac{(p+1) (D-p-3)}{D-2}  + \frac{4 (D-p-2) 
(D-2p -4)}{D-2}\nonumber \\
&& + [ 4(D-p-2) -2] +  [2 - \frac{(p+1)(D-p-3)}{D-2}]    \Bigg\} \;,
\label{chiTOTALEFINALE}
\end{eqnarray}
where $\tau_p$ is the physical D$p$-brane tension, $\tau_p = \frac{T_p}{\kappa_D}$. The first line contains the leading contribution given by the tree diagram with a graviton exchange and the subleading term of one-loop diagram with two graviton exchanges, the third line gives  the contribution of the one-loop seagull diagram and the fourth line gives the contribution of the one-loop diagram with the 3-graviton vertex. Finally the first square bracket in the last line gives the contribution of the one-loop diagram with the RR fields attached to the D$p$-branes, while the last square bracket gives that of the dilaton attached to the D$p$-branes. 

It is easy to show, in this extended theory of gravity, that the subleading contribution contained inside the big curly brackets vanishes. In this case the sum of the leading and subleading eikonal reduces just to the expression in the first line of (\ref{chiTOTALEFINALE}).  This is in agreement with the same result obtained in \cite{D'Appollonio:2010ae} for $D=10$ and the results found in section \ref{HElimit}.

\subsection{Pure Einstein Gravity}
\label{Einstein}

In this section we will consider the case of pure Einstein gravity. Let us start by considering the one-point graviton amplitude where only the tree diagram with the graviton exchange and the
one-loop diagram with the three-graviton vertex contribute. They are given in momentum space by (\ref{gra13}) and (\ref{3gra1}), respectively. Going to impact parameter space we find,
\begin{eqnarray}
&&\langle \eta_{\mu \nu} + 2 \kappa_D h_{\mu \nu} \rangle \nonumber \\
&&  = \left[ 1- \frac{D-p-3}{D-2} \left( \frac{R_p}{r}\right)^{D-p-3} + \frac{1}{2}
\left( \frac{D-p-3}{D-2} \right)^2 \left( \frac{R_p}{r}\right)^{2(D-p-3)}+ \dots  \right]\eta_{\parallel \mu \nu}
\nonumber \\
&& + \left[ 1 + \frac{p+1}{D-2}  \left( \frac{R_p}{r}\right)^{D-p-3}  \right. \nonumber \\
&& \left.  - \frac{1}{4} 
\left( \frac{  (D-p-3)^2 (p+1) }{2 (D-2) (D-p-2)} - 2 \left( \frac{p+1}{D-2}  \right)^2 \right) 
\left( \frac{R_p}{r}\right)^{2(D-p-3)}+ \dots  \right] \eta_{\perp \mu \nu} \nonumber \\
&& - \frac{1}{4 (5+p-D)} 
\left( \frac{(D-p-3)^2 (p+1)}{2 (D-2) (D-p-2)} - \frac{2(p+1) (D-p-3)}{D-2} \right) \nonumber \\
&& \times
\left( \eta_{\perp \mu \nu} - 2(D-p-3) \frac{r_\mu r_\nu}{r^2} \right)  \left( \frac{R_p}{r} \right)^{2(D-p-3)} \;, \label{EDS11}
\end{eqnarray}
where in the right-hand-side we have added the contribution of the flat Minkowski metric for $r \rightarrow \infty$. Notice that in the equation above we have now included the term in the third line of (\ref{3gra1}) that was neglected  in  subsection \ref{1point} and the term  in the second line of the same equation that was cancelled by the additional contributions of the dilaton and RR field. It can be checked that the term in the third line of (\ref{3gra1}), that we have neglected, gives the second term in the second to last line of (\ref{EDS11}).

To make contact with existing literature let us consider the case $D=4$ and $p=0$ where
\begin{eqnarray}
N \tau_0 = \frac{N T_0}{\kappa_4}  \equiv M ~~~;~~~R_p \rightarrow 4 G_N M \;.
\label{EDS12}
\end{eqnarray}
Then (\ref{EDS11}) becomes
\begin{eqnarray}
&&\langle \eta_{\mu \nu} + 2 \kappa_4 h_{\mu \nu} \rangle =  \left[ 1 - \frac{4MG_N}{2 r} + \frac{1}{8} \left( \frac{4MG_N}{r}\right)^2   + \dots \right]\eta_{00} \nonumber \\
&& + \left[ 1 + \frac{4MG_N}{2 r} + \frac{1}{4}(\frac{1}{4} +1 ) \left( \frac{4MG_N}{r}\right)^2   \right] \eta_{ij} + ( \frac{1}{8} -1) \frac{r_i r_j}{2 r^2}  \left( \frac{4MG_N}{r}\right)^2   \;,
\label{EDS13}
\end{eqnarray}
where the second term in the two round brackets in the last line comes from the term in the third line of (\ref{3gra1}). The subscript $0$ corresponds to the time coordinate, while $i,j$ correspond to the three spatial coordinates.  It is easy to check that the previous metric satisfies the following condition at each order in $G_N$,
\begin{eqnarray}
\partial^\nu h_{\nu \mu} - \frac{1}{2} \partial_\mu h =0 ~~;~~~ h \equiv h_{\mu \nu} \eta^{\mu \nu} \;.
\label{EDS13a}
\end{eqnarray}
If we want the one-point amplitude in the harmonic gauge the term of order $G_N^2$ must satisfy (54) of \cite{BjerrumBohr:2002ks} instead of the equation above. This is obtained by neglecting in (\ref{EDS13}) the second term in the two round brackets.  With this gauge choice (\ref{EDS13}) becomes,
\begin{eqnarray}
\langle g_{\mu \nu} \rangle = \left( 1 - \frac{2MG_N}{r} + \frac{2 M^2 G_N^2}{r^2} \right) \eta_{00} +
\left( 1 + \frac{2MG_N}{r} + \frac{M^2 G_N^2}{r^2} \right) \eta_{ij} + \frac{r_i r_j}{r^2} \frac{M^2 G_N^2}{r^2}
\label{EDS13b}
\end{eqnarray}


In the final part of this subsection we consider the leading and subleading eikonal in the case of pure Einstein gravity. It can be easily obtained from the one in  (\ref{chiTOTALEFINALE}) by neglecting the last line. We find that,
\begin{eqnarray}
&& \chi (b, E) = \frac{N \kappa_D^2 \tau_p}{4}  E\frac{\Gamma 
( \frac{D-p-4}{2})}{ \pi^{\frac{D-p-2}{2}} b^{D-p-4}} 
+\frac{(N \kappa_D^2 \tau_p)^2 \, E \,\Gamma^2 ( \frac{D-p-3}{2}) 
\Gamma (D-p - \frac{7}{2})}{16 \,\pi^{D-p - \frac{3}{2}} \Gamma (D-p-4) 
\,\,b^{2D-2p-7}} \nonumber \\
&& + \frac{(N \kappa_D^2 \tau_p)^2 \, E \, \Gamma (D-p - \frac{7}{2})
\Gamma^2 ( \frac{D-p-1}{2})}{16 (3+p-D) \Gamma (D-p-1)  
\,\,\pi^{D-p - \frac{3}{2}}\,\, b^{2D-2p-7}}   \nonumber \\
&& \times  
\Bigg\{ - 4 (D-p-2) +  \frac{(p+1) (D-p-3)}{D-2} \Bigg\} \;.
\label{chiEINSTEIN}
\end{eqnarray}
If we look at the case for $D=4$ and $p=0$, we see that the last term in the first line does not contribute and  regularising the first term,
\begin{eqnarray}
\frac{\Gamma (\frac{D-p-4}{2})}{b^{D-p-4}} \Longrightarrow  -2 \log b \;,
\label{EDS14}
\end{eqnarray}
we find that,
\begin{eqnarray}
&&\chi^{(D=4; p=0)} = - 4 G_N ME \log b + \frac{\pi (G_N M)^2 E}{2b} \left(8 - \frac{1}{2} \right) \nonumber \\
&& =  - 4  G_N ME \log b +  \frac{15 \pi (G_N M)^2 E}{4b} \;,
\label{EDS15}
\end{eqnarray}
where the first term in the round bracket comes from the seagull diagram, while the second comes from the one-loop diagram with the 3-graviton vertex. It agrees with the classical part of the eikonal derived in \cite{Bjerrum-Bohr:2016hpa} and with the eikonal derived in \cite{Akhoury:2013yua,Luna:2016idw}. From the eikonal we can derive the deflection angle for a massless particle,
\begin{eqnarray}
\theta = - \frac{1}{E} \frac{\partial}{\partial b} \chi^{(D=4; p=0)} = \frac{4 G_N M}{b} + \frac{15 \pi (G_N M)^2}{4b^2} + \dots
\label{EDS16}
\end{eqnarray}
where the dots refer to terms with higher powers of $b$ in the denominator. The first term is the old result from Einstein, while the second term agrees with recent calculations in \cite{Akhoury:2013yua,Bjerrum-Bohr:2016hpa,Luna:2016idw, Bjerrum-Bohr:2018xdl}. 

Using \eqref{chiEINSTEIN} we can also calculate the deflection angle for $D$ dimensions and $p=0$. We find,

\begin{eqnarray}
\theta = - \frac{1}{E} \frac{\partial}{\partial b} \chi^{(p=0)} = \sqrt{\pi}\frac{\Gamma \left( \frac{D}{2} \right)}{\Gamma \left( \frac{D-1}{2} \right)} \left( \frac{R_s}{b}\right)^{D-3} + \frac{\sqrt{\pi}}{2} \frac{\Gamma \left(D-\frac{1}{2} \right)}{\Gamma \left(D-2 \right)}\left( \frac{R_s}{b}\right)^{2D-6} \;,
\end{eqnarray}
where $R_s$ is the ``Schwarzschild radius" defined in appendix \ref{Ddimdeflection}. Comparing this result with \eqref{eq:deflect5}, where the deflection angle has been calculated from the metric for the D-dimensional generalisation of a Schwarzschild black hole, we find perfect agreement. Note that we cannot compare the result for general $p$ because the D-brane coupling used in \cite{Emparan:2009at} is different to the one we are using here.


\section{Discussion}
\label{sec:discussion}

We have discussed how to extract, from scattering amplitudes, classical quantities such as the classical solution related to the backreaction of a heavy source and the eikonal describing a scattering process in the Regge regime. The general ideas are well known and have been exploited in several previous papers to obtain these quantities in the limit of large distance or impact parameter, see for instance~\cite{Duff:1974xx,Bertolini:2000jy}. In this paper we have presented a detailed analysis of the first subleading corrections to the limit mentioned above by focusing on type II supergravity in the presence of a stack of parallel D$p$-branes as an example. In the case of the eikonal, these corrections are determined by the subleading energy contributions in the Regge regime and so probe the structure of the gravitational theory in more detail. For instance the leading eikonal receives contributions only from ladder diagrams where gravitons are exchanged, while the subleading eikonal involves diagrams with different topologies and lower spin states. This raises the possibility, at the first subleading order, that the eikonal should be described by an operator instead of a simple phase since inelastic processes become possible\footnote{In gravitational theories with a higher derivative modification of the 3-graviton vertex this happens already at the level of leading eikonal~\cite{Camanho:2014apa}.}.

We studied this possibility in the context of type II supergravity focusing on the scattering of massless perturbative states from a stack of D$p$-branes. The two main points of our analysis for the subleading eikonal are that the relevant information is encoded in the onshell three and four-point vertices (see section~\ref{2braneamps}) and that its derivation requires us to disentangle cross terms between leading and subleading energy contributions (see section~\ref{HElimit}). In the case under study, the inelastic contributions which grow with energy in the one-loop amplitudes are completely accounted for by the cross terms mentioned above and thus the final expression for the first subleading eikonal in supergravity is described fully by the elastic processes and is given by,
\begin{equation}
\chi^{(2)}(E,b) = \frac{(N T_p \kappa_D)^2 E}{16 \pi^{D-p-\frac{3}{2}}} \frac{\Gamma^2 \left(\frac{D-p-3}{2} \right) \Gamma \left(\frac{2D-2p-7}{2} \right)}{\Gamma \left(D-p-4 \right)} \frac{1}{b^{2D-2p-7}}    \;,
\end{equation}
which agrees with~\cite{D'Appollonio:2010ae}. We do not know a general argument proving that this is always the case and so we think that it would be interesting to check this pattern both in more complicated theories and at further subleading orders. For instance, an analysis of the eikonal operator in string theory beyond the leading order~\cite{Amati:1987wq,Amati:1987uf,Amati:1988tn} is missing. Of course we could use the full four-point string amplitudes in our derivation of section~\ref{2braneamps} simply by reinstating the $\alpha'$ dependence that in maximally supersymmetric theories appears just in the overall combination of gamma functions,
\begin{equation}
  \label{eq:gammafst}
   \frac{\Gamma ( 1 -\frac{\alpha' s}{4} ) \Gamma ( 1- \frac{\alpha't}{4} )
\Gamma ( 1 -\frac{\alpha'u}{4} ) }{\Gamma ( 1  + \frac{\alpha's}{4}) \Gamma ( 1 +\frac{\alpha't}{4} ) \Gamma ( 1 +\frac{\alpha'u}{4} )}\;.
\end{equation}
For instance in the first contribution to the dilaton to dilaton scattering we analysed, this amounts to including the factor defined above in the vertex~\eqref{STampRR}. However this is not sufficient to reconstruct the full string eikonal as, of course, we need to include also the contributions due the exchanges of the leading and subleading Regge trajectories between the D$p$-branes and the perturbative states. It seems possible to generalise, along these lines, the analysis of this paper to the full string setup by using the formalism of the Reggeon vertex~\cite{Ademollo:1989ag,Ademollo:1990sd,Brower:2006ea, D'Appollonio:2013hja}. 

A parallel development, entirely within field theory, is to analyse further subleading contributions to the eikonal. As mentioned in the introduction, there is a practical motivations for doing so since from these results it is possible to extract new information on the one body effective action which is  used in the analysis of gravitational waves. Of course, in this context, the interesting setup is that of $2\to 2$ scattering with objects with large but finite masses, $m_1$ and $m_2$. The approach discussed here could also be applied in this case. A further interesting generalisation is to include a non-zero angular momentum for the external massive states, so that they can represent spinning black holes, and extract information on the one body effective action in this case~\cite{Bini:2017xzy,Vines:2017hyw}. On the more conceptual side, it would be interesting to check if in these more general cases the subleading contributions to the eikonal are still universal or if there are inelastic effects that induce differences between the various states as we know happens when the 3-graviton vertex is modified~\cite{Camanho:2014apa}.

\section*{Acknowledgements}

We would like to thank  E. Bjerrum-Bohr, R. Marotta and G. Veneziano for discussions and correspondence. This work was supported in part by the Science and Technology Facilities Council (STFC) Consolidated Grant ST/L000415/1 {\it String theory, gauge theory \& duality}. AKC is supported by an STFC studentship.


\appendix
\section{Feynman Rules} \label{appFeynRules}

Here we will outline the Feynman rules that we have used throughout this paper. We will neglect writing the various momentum conserving delta functions associated with the various vertices. We take the dilaton as $\phi$, the graviton as $h$ and the RR gauge field as $C^{(n-1)}$, where $n=p+2$ and $p$ is the dimension of the D$p$-brane world-volume. We start by writing the bulk action that we will use to derive the Feynman rules,
\begin{eqnarray}\label{eq:bulkaction}
S= \int d^{D} x \sqrt{-g} \left[ \frac{1}{2 \kappa_D^2} R- \frac{1}{2} 
\partial_{\mu}\phi\, \partial^{\mu}\phi -
\frac{1}{2n!}e^{-a(D) \sqrt{2}\kappa_D \phi }F^2_{n} \right] \text{ ,}
\end{eqnarray}
where the expression for $a(D)$ can be found in \cite{Duff:1994an} and is given by\footnote{In 10D type II supergravity, $a(D=10)=\frac{p-3}{2}$.},
\begin{equation}
a^2(D)=4-\frac{2(p+1)(D-p-3)}{D-2} \;.
\end{equation}
Note that by convention we will use the positive root of the expression above. We will use the following boundary action which is being sourced by the D$p$-branes,
\begin{eqnarray}
\int d^D x \,\,\delta^{D-p-1} (x) \left( J_h h_{\alpha}^{\,\,\, \alpha} (x)+
J_{\phi} \phi (x) + \mu_p C_{01 \dots p} (x) \right)
\label{branecoup}
\end{eqnarray}
where the quantities $J_h, J_{\phi}, \mu_p$ are the couplings of the graviton, dilaton 
and RR to the D$p$-brane. They are given by
\begin{eqnarray}
J_h = - T_p~~~;~~~~
\mu_p = \sqrt{2} T_p  \,\,;\,\,\,\, J_\phi= - \frac{a(D)}{\sqrt{2}}T_p \;.
\label{JJmu}
\end{eqnarray}
We expand the metric as,
\begin{eqnarray}
g_{\mu\nu}= \eta_{\mu\nu} + 2 \kappa_{D} h_{\mu\nu} 
\end{eqnarray}
and from the quadratic part of the action~\eqref{eq:bulkaction} plus the de Donder gauge fixing term, we obtain the following graviton propagator of momentum $q$
\begin{equation}\label{eq:dedop}
[G_{h}]^{\mu \nu ; \rho \sigma} = \frac{-i}{2q^2} \left ( \eta^{\mu \rho} \eta^{\nu \sigma} + \eta^{\mu \sigma} \eta^{\nu \rho} - \frac{2}{D-2} \eta^{\mu \nu} \eta^{\rho \sigma} \right ) \text{ .}
\end{equation} 
Similarly, from the RR kinetic term and the Feynman gauge fixing term we obtain the RR field propagator
\begin{equation}
[G_{C^{(n-1)}}]^{\mu_1\ldots\mu_{n-1}}_{\nu_1\ldots\nu_{n-1}} = \frac{-i}{q^2}\,(n-1)! \delta^{\mu_1}_{[\nu_1} \ldots \delta^{\mu_{n-1}}_{\nu_{n-1}]} \text{ .}
\end{equation}
Finally we have the standard scalar propagator for the dilaton
\begin{equation}
[G_{\phi}] = \frac{-i}{k^2} \text{ .}
\end{equation}
The couplings that are relevant for the scattering process involving RR fields as sources are the $\phi^2 h$ 
\begin{equation}
[V_{\phi_1 \phi_2 h}] = -i \kappa_D \left (k_{1\, \mu}k_{2\,\nu} + k_{1\,\nu}k_{2\, \mu} - k_1 \cdot k_2 \eta_{\mu \nu} \right ) h^{\mu \nu} \text{ ,} \label{ddgV}
\end{equation}
where $k_1$ and $k_2$ are the momenta associated with the dilatons. The $\phi \; C^{(n-1)} C^{(n-1)}$ vertex is given by,
\begin{equation}
[V_{\phi F^{(n)}_{1} C_2^{(n-1)}}] = - \frac{i a(D) \sqrt{2} \kappa_D}{(n-1)!} (F_{1\; \mu_1 \mu_2 \ldots \mu_n} k^{\mu_1}_{2} C_2^{\mu_2 \ldots \mu_n}) \text{ ,}
\end{equation}
where (1) and (2) are labels of the two RR fields and $F_{i\; \mu_1 \ldots \mu_n}$ is the field strength associated with the $(p+1)$-form gauge field $C_{i \; \mu_2 \ldots \mu_n}$, $F_{i\; \mu_1 \ldots \mu_n}=(\text{d}C_{i})_{\mu_1 \ldots \mu_n}$. The $h \; C^{(n-1)} C^{(n-1)}$ vertex is given by,
\begin{equation}
[V_{F^{(n)}_{1} F^{(n)}_{2} h}] = \frac{i \kappa_D}{n!} (2n F_{12}^{\mu \nu} - \eta^{\mu \nu} F_{12}) h^{\mu \nu} \mbox{ ,}
\end{equation}
where $F_{12}^{\mu \nu} = F^{\mu}_{1 \; \mu_2 \ldots \mu_n} F_{2}^{\nu \mu_2 \ldots \mu_n}$ and we also have $F_{12} = F_{12}^{\mu \nu} \eta_{\mu \nu}$. The $\phi^2 \; C^{(n-1)} C^{(n-1)}$ vertex is given by,
\begin{equation}
[V_{\phi \phi' F^{(n)}_{1} F^{(n)}_{2}}] = - \frac{2 i \kappa^2}{n!} 
a^2(D) 
F_{12} \mbox{ .}
\end{equation}
The extra Feynman rules that are relevant for when considering graviton sources are shown below. The $h^3$ vertex is given by, 
\begin{eqnarray}
&&[V_{h_{1k} h_{2p} h_{3q}}] 
(k,p,q) = - 2 i \kappa_D \bigl( - \frac{1}{2} p_{\mu} q_{\nu} \eta_{\lambda \rho} 
\eta_{\tau \sigma} + 2 p_{\mu} q_{\sigma } \eta_{\lambda \nu} \eta_{\tau \rho} - \eta_{\rho \sigma} p_{\mu} q_{\lambda} \eta_{\tau \nu}- \frac{1}{2} \eta_{\mu \nu} p_{\tau} \eta_{\lambda \rho} q_{\sigma}
\nonumber \\
&& + \frac{1}{4} \eta_{\mu \nu} \eta_{\lambda \rho} \eta_{\tau \sigma } p\cdot q  - \eta_{\rho \sigma}
p_{\lambda} \eta_{\mu \tau} q_{\nu} + \frac{1}{2} \eta_{\rho \sigma} p_{\mu} 
\eta_{\lambda \tau} q_{\nu}  - \eta_{\mu \rho} \eta_{\nu \sigma} p_{\lambda} q_{\tau}
+ \frac{1}{2} \eta_{\mu \nu} \eta_{\rho \sigma} p_{\lambda} q_{\tau}
\\ \nonumber 
&& + \eta_{\mu \rho} \eta_{\nu \sigma} \eta_{\lambda \tau} p \cdot q  - \frac{1}{4} \eta_{\mu \nu} \eta_{\rho \sigma}  \eta_{\lambda \tau} p \cdot q - \eta_{\mu \sigma} \eta_{\nu \lambda} \eta_{\rho \tau} p \cdot q + \eta_{\mu \sigma} p_{\lambda} \eta_{\nu \tau} q_{\rho} \bigr) h_1^{\mu \nu} h_2^{\rho\sigma} h_3^{\lambda \tau}  + \mbox{\ldots ,}
\label{3graV}
\end{eqnarray}
where the dots stand for the sum over the permutations of the external states and, as usual,  $k$, $p$ and $q$ indicates their momenta. The $\phi^2 h^2$ vertex is given by,
\begin{equation}
[V_{\phi_1 \phi_2 h_3 h_4}] 
= (i \kappa_D^2) \left[ k_1 k_2 \left(\frac{1}{2} \eta_{\rho \sigma}\eta_{\lambda \tau} - \eta_{\rho \tau} \eta_{\sigma \lambda} \right) + 4 k_{1\rho } k_{2\tau} \eta_{\sigma \lambda} - 2 k_{1\rho} k_{2 \sigma}\eta_{\lambda \tau} \right] h_3^{\rho\sigma}  h_4^{\lambda \tau} \text{+ ... ,}
\end{equation}
where the dots stand for the symmetrisation between the two dilatons and the two gravitons, while $k_1$ and $k_2$ are the momenta of the dilatons. From the Born-Infeld boundary action we read the D$p$-brane graviton coupling
\begin{equation}
[B_{h}] = -i T_p \eta^{\mu \nu}_{\parallel} h_{\mu \nu} \int \frac{d^{\perp}k}{(2 \pi)^{\perp}} \text{ ,} \label{branecouplinggrav}
\end{equation}
where $\parallel$ denotes the $p+1$ directions along the D$p$-brane and $\perp$ denotes the $D-p-1$ directions transverse to the D$p$-brane. The boundary coupling with the dilaton is given by
\begin{equation}
[B_{\phi}] = - i T_p \frac{a(D)}{\sqrt{2}} \int \frac{d^{\perp}k}{(2 \pi)^{\perp}} 
\end{equation}
and the coupling with RR gauge potential is given by,
\begin{equation}
[B_{C^{(n-1)}}] = i \mu_p C_{01\ldots p} \int \frac{d^{\perp}k}{(2 \pi)^{\perp}} \text{ .}
\end{equation}

\section{Integral Reference} \label{integralref}

We define an $m$-index, $p$-propagator integral as,

\begin{equation}
\mathcal{I}_{p}^{\mu_1 \ldots \mu_m} = \int \frac{d^{\perp} k}{(2\pi)^{\perp}} \frac{k^{\mu_1} \ldots k^{\mu_m}}{k^2 (k-l_1)_{\perp}^2 \ldots (k-l_{p-1})_{\perp}^2} \text{ ,}
\end{equation}
where the variables $l$ contain only external momenta and $\perp=D-p-1$. We give some explicit results below which are useful when considering high-energy contributions.

\subsection{3-Propagator Scalar Integral}

For the scalar integral with three propagators we have,

\begin{equation}
\mathcal{I}_{3}(q_{\perp}) = \int \frac{d^{\perp} k}{(2\pi)^{\perp}}\frac{1}{k^2 (k_1-k)_{\perp}^2 (k+k_2)_{\perp}^2} = \mathcal{I}_{3}^{(1)}(q_{\perp}) + \mathcal{I}_{3}^{(2)}(q_{\perp}) + \ldots \;, \label{I3full}
\end{equation}
where $k^2= k_{\perp}^2 - E^2$ and $\mathcal{I}_{3}^{(1)}(q_{\perp})$ is the leading energy contribution given by,
\begin{equation}
\mathcal{I}_{3}^{(1)}(q_{\perp}) = \frac{\sqrt{\pi}}{(4\pi)^{\frac{D-p-1}{2}}} (q^2_{\perp})^{\frac{D-p-6}{2}}  \frac{i}{2 E} \frac{\Gamma{\left(\frac{6-D+p}{2} \right)} \Gamma^2(\frac{D-p-4}{2})}{\Gamma(D-p-4)} \text{ ,}
\end{equation}
and $\mathcal{I}_{3}^{(2)}(q_{\perp})$ is the subleading contribution,
\begin{equation}
\mathcal{I}_{3}^{(2)}(q_{\perp}) = - \frac{1}{(4\pi)^{\frac{D-p-1}{2}}} (q^2_{\perp})^{\frac{D-p-5}{2}}  \frac{1}{2 E^2} \frac{\Gamma{\left(\frac{5-D+p}{2} \right)} \Gamma^2(\frac{D-p-3}{2})}{\Gamma(D-p-4)} \text{ .}
\end{equation}
We briefly review what happens when looking at $\mathcal{I}_3$, if we remove the $k+k_2$ propagator in \eqref{I3full} to explicitly recognise that these contributions are localised on the D-branes when working in impact parameter space. We find after introducing Schwinger parameters,

\begin{equation}
\mathcal{I}_{3,3} = \int \frac{d^{\perp} k}{(2\pi)^{\perp}}\frac{1}{k^2 (k_1-k)_{\perp}^2} =  \int dt_1 dt_2 \left( \frac{\pi}{T} \right)^{\frac{\perp}{2}} \text{exp} \left[ \frac{t_2^2 k_{1 \perp}^2}{T} +t_1^2 E^2 - t_2 k_{1 \perp}^2 \right] 
\end{equation}
where $\mathcal{I}_{3,3}$ refers to the fact that we've killed the third propagator in the integral and $T=t_1+t_2$ where $t_1$, $t_2$ are the Schwinger parameters. From this we can see that $\mathcal{I}_{3,3}=f(E)$ which means that the result is not a function of the momentum exchanged, $q_{\perp}$. If we calculate the impact parameter space expression for this integral we find that $\mathcal{\tilde{I}}_{3,3}= f(E) \delta^{\perp-1}(\mathbf{b})$, which as described before suggests that these types of terms can only produce contributions which are localised on the D-branes. Note that the same happens when you remove the second propagator $k_1-k$. However if one removes the propagator $k$ we find that $\mathcal{I}_{3,1}=\mathcal{I}_{2}$ as expected.

\subsection{2-Propagator Integrals} \label{integralref2prop}

For the scalar integral with two propagators we have,

\begin{equation}
\mathcal{I}_{2}(q_{\perp}) = \int \frac{d^{\perp} k}{(2\pi)^{\perp}} \frac{1}{k_{\perp}^2 (k-q)_{\perp}^2} = \frac{1}{(4\pi)^{\frac{D-p-1}{2}}} (q^2_{\perp})^{\frac{D-p-5}{2}}  \frac{\Gamma{\left( \frac{3-D+p}{2} \right)} \Gamma^2{\left( \frac{D-p-1}{2} \right)}}{ \Gamma{\left( D-p-1 \right)}} (-2(D-p-2)) \text{ .}
\end{equation}
For the 1-index integral with two propagators we have,

\begin{equation}
\mathcal{I}_{2}^{\mu}(q_{\perp}) = \int \frac{d^{\perp} k}{(2\pi)^{\perp}} \frac{k^{\mu}}{k_{\perp}^2 (k-q)_{\perp}^2} = \frac{1}{(4\pi)^{\frac{D-p-1}{2}}} (q^2_{\perp})^{\frac{D-p-5}{2}} q^{\mu} \frac{\Gamma{\left( \frac{5-D+p}{2} \right)} \Gamma{\left( \frac{D-p-1}{2} \right)} \Gamma{\left( \frac{D-p-3}{2} \right)}}{\Gamma{\left( D-p-2 \right)}} \text{ .}
\end{equation}
For the 2-index integral with two propagators we have,

\begin{eqnarray}
\mathcal{I}_{2}^{\mu \nu}(q_{\perp}) = \int \frac{d^{\perp} k}{(2\pi)^{\perp}} \frac{k^{\mu}k^{\nu}}{k_{\perp}^2 (k-q)_{\perp}^2} &=& \frac{1}{(4\pi)^{\frac{D-p-1}{2}}} (q^2_{\perp})^{\frac{D-p-3}{2}}  \frac{\Gamma{\left( \frac{3-D+p}{2} \right)} \Gamma^2{\left( \frac{D-p-1}{2} \right)}}{2 \Gamma{\left( D-p-1 \right)}} \bigl( \eta^{\mu \nu}_{\perp}  \nonumber \\ 
&& - (D-p-1) \frac{q^{\mu}q^{\nu}}{q^2} \bigl) \text{ .}
\end{eqnarray}

\subsection{Other Integrals}

Here we list some of the other integrals that we have used throughout the paper. In order to calculate the impact parameter space expressions in sections \ref{HElimit} and \ref{lntoeikonal} we have used, 

\begin{eqnarray}
\int \frac{d^D k}{(2\pi)^D} {\rm e}^{i k \cdot b} (k^2)^\nu = \frac{2^{2\nu}}{\pi^{D/2}}
\frac{\Gamma ( \nu + \frac{D}{2})}{\Gamma (-\nu)} 
\frac{1}{ (b^2)^{\nu + \frac{D}{2}}} \;. \label{impoformu}
\end{eqnarray}

\section{Relations for Manipulating RR Field Strengths} \label{RRidentities}

In this appendix we will explicitly analyse the various types of products between field strengths and momenta that arise in \eqref{RRRRdgrawamp}. It is important to recall that in these expressions one of the RR fields is attached to the D-branes (with label 4) and one is an external state (with label 2). We will first focus on the term that is relevant for the leading energy contribution and subsequently look at all other combinations. We have,

\begin{eqnarray}
E^2 F_{24} &=& E^2 \left( k_2^{\mu_1} C^{(2) \mu_2 \ldots \mu_{n}} + (-1)^{n-1} k_2^{\mu_2} C^{(2) \mu_3 \ldots \mu_{n} \mu_1} + \ldots \right) \nonumber \\
&& \times \left[ k_{4 \mu_1} C^{(4)}_{\mu_2 \ldots \mu_{n}} + (-1)^{n-1} k_{4 \mu_2} C^{(4)}_{\mu_3 \ldots \mu_{n} \mu_1} + \ldots \right) \\ \nonumber
&=& E^2 \left( n (k_2 \cdot k_4) C^{(2) \mu_2 \ldots \mu_{n}} C^{(4)}_{\mu_2 \ldots \mu_{n}} + (-1)^{n-1} n(n-1) k_2^{\mu_1} C^{(2) \mu_2 \ldots \mu_{n}} k_{4 \mu_2} C^{(4)}_{\mu_3 \ldots \mu_{n} \mu_1} \right] \text{ ,}
\end{eqnarray}
where we have identified the two terms in the last line above to be the only two distinct types of terms that arise, with the associated counting taken into account. The indices on $C^{(4)}$ have to lie in the space parallel to the D-brane world volume so $\mu_3 \ldots \mu_{n} = 1 \ldots p$ and we note that there are $(n-2)!$ ways to do this. Furthermore $(k_2^{\mu_1})_{\parallel}$ only has a non-zero component for $\mu_1 = 0$ ($\mu_1$ has to also lie along the D-branes as it is one of the indices on $C^{(4)}$) and so we have to all orders in $E$,

\begin{eqnarray}
E^2 F_{24} &=& E^3 n! (k_4 \cdot C^{(2)})^{1 \ldots p} C^{(4)}_{0 \ldots p} + E^2 n! (k_2 \cdot k_4) C^{(2) 0 \ldots p} C^{(4)}_{0 \ldots p}
\end{eqnarray}
where we have used $(k_{2})^{0}=-E$ and $(-1)^{n-1} C^{(4)}_{1 \ldots p 0} = - C^{(4)}_{0 \ldots p}$. 

We now look at types of terms which contribute to the subleading energy behaviour of \eqref{RRRRdgrawamp},

\begin{eqnarray}
F_{24}^{\alpha \mu} \eta_{\parallel \alpha \nu}k_1^{\nu}k_{3 \mu} &=&  F_{24}^{0 \mu} (k_{1})_{0} k_{3 \mu} \nonumber \\
&=&  \left( (k_2)^{0} C^{(2) \mu_2 \ldots \mu_{n}} + (-1)^{n-1} k_2^{\mu_2} C^{(2) \mu_3 \ldots \mu_{n} 0} + \ldots \right) \nonumber \\
&& \times \left( k_{4}^{\mu} C^{(4)}_{\mu_2 \ldots \mu_{n}} + (-1)^{n-1} k_{4 \mu_2} C^{(4)}_{\mu_3 \ldots \mu_{n}}{}^{\mu} + \ldots \right) (k_{1})_{0} k_{3 \mu} \nonumber \\
&=& \left( (k_2)^{0} C^{(2) \mu_2 \ldots \mu_{n}} + (-1)^{n-1} k_2^{\mu_2} C^{(2) \mu_3 \ldots \mu_{n} 0} + \ldots \right) \left( k_{4}^{\mu} C^{(4)}_{\mu_2 \ldots \mu_{n}} \right) (k_{1})_{0} k_{3 \mu} \nonumber \\
&=&  (n-1)! \left( (k_2)^{0} C^{(2) 0 \ldots p} + (-1)^{n-1} k_2^{0} C^{(2) 1 \ldots p 0} \right) \left( (k_3 \cdot k_{4}) C^{(4)}_{0 \ldots p} \right) (k_{1})_{0}    \nonumber \\
&=& 0 \text{ ,}
\end{eqnarray}
where we have used the fact that the $\mu$ index has to be along the D-branes but $(k_3)_{\parallel}=0$ in the third line and in the last line we have again used the fact that $(-1)^{n-1} C^{(2)}_{1 \ldots p 0} = - C^{(2)}_{0 \ldots p}$. We also need,
\begin{eqnarray}
F_{24}^{\alpha \beta} \eta_{\parallel \beta \mu} k_1^{\mu} &=& F_{2}^{\alpha \mu_2 \ldots \mu_n} \left( k_{4}^{\beta} C^{(4)}_{\mu_2 \ldots \mu_{n}} + (-1)^{n-1} k_{4 \mu_2} C^{(4)}_{\mu_3 \ldots \mu_{n}}{}^{\beta} + \ldots \right) \eta_{\parallel \beta \mu} k_1^{\mu} \nonumber \\
&=& - (n-1)!  F_{2}^{\alpha \mu_2 1 \ldots p} k_{4 \mu_2} (k_1)^{0} C^{(4)}_{0 \ldots p} \nonumber \\
&=& - (n-1)! E \left[( k_4 \cdot C^{(2)})^{1 \ldots p} k_2^{\alpha} - (k_2 \cdot k_4) C^{(2) \alpha 1 \ldots p} \right] C^{(4)}_{0 \ldots p} \;,
\end{eqnarray}
where in the second line we have used the fact that $(k_1 \cdot k_4)_{\parallel} = 0$. We also require,

\begin{eqnarray}
F_{24}^{\alpha \beta} k_{1 \beta} &=& F_{2}^{\alpha \mu_2 \ldots \mu_n} \left( k_{4}^{\beta} C^{(4)}_{\mu_2 \ldots \mu_{n}} + (-1)^{n-1} k_{4 \mu_2} C^{(4)}_{\mu_3 \ldots \mu_{n}}{}^{\beta} + \ldots \right) k_{1 \beta} \nonumber \\
&=& (n-1)! \left( (k_1 \cdot k_{4}) F_{2}^{\alpha 0 \ldots p} C^{(4)}_{0 \ldots p} - k_{4 \mu_2} (k_1)^{0} F_{2}^{\alpha \mu_2 1 \ldots p} C^{(4)}_{0 \ldots p} + \ldots \right)\nonumber \\
&=& (n-1)! \bigl[ (k_1 \cdot k_4) \left(k_2^{\alpha} C^{(2) 0 \ldots p} + E C^{(2) \alpha 1 \ldots p} \right) \nonumber \\
&& - E ( k_4 \cdot C^{(2)})^{1 \ldots p} k_2^{\alpha} + E (k_2 \cdot k_4) C^{(2) \alpha 1 \ldots p} \bigr] C^{(4)}_{0 \ldots p} \;,
\end{eqnarray}
where we have used the properties of the gauge potentials outlined previously. We can also have,

\begin{eqnarray}
F_{24}^{\alpha \beta} k_{3 \beta} &=& F_{2}^{\alpha \mu_2 \ldots \mu_n} \left( k_{4}^{\beta} C^{(4)}_{\mu_2 \ldots \mu_{n}} + (-1)^{n-1} k_{4 \mu_2} C^{(4)}_{\mu_3 \ldots \mu_{n}}{}^{\beta} + \ldots \right) k_{3 \beta} \nonumber \\
&=& (n-1)! \left( (k_3 \cdot k_{4}) F_{2}^{\alpha 0 \ldots p} C^{(4)}_{0 \ldots p} \right)\nonumber \\
&=& (n-1)! (k_3 \cdot k_4) \left(k_2^{\alpha} C^{(2) 0 \ldots p} + E C^{(2) \alpha 1 \ldots p} \right) C^{(4)}_{0 \ldots p} \;,
\end{eqnarray}
where in the second line we have again used the fact that the index $\beta$ must lie parallel to the D-branes but $(k_3)_{\parallel}=0$. Finally we have,

\begin{eqnarray}
F_{24}^{\alpha \beta \mu \nu} \eta_{\parallel \beta \nu} k_{3 \alpha} k_{3 \mu} &=& F_{2}^{\alpha \beta \mu_3 \ldots \mu_n} \bigl( k_{4}^{\mu} C^{(4) \nu}{}_{\mu_3 \ldots \mu_{n}} + (-1)^{n-1} k_4^{\nu} C^{(4)}_{\mu_3 \ldots \mu_{n}}{}^{\mu} + (-1)^{n-1} k_{4 \mu_3} C^{(4)}_{\mu_4 \ldots \mu_{n}}{}^{\mu \nu} \nonumber \\
&& + \ldots \bigr) \eta_{\parallel \beta \nu} k_{3 \alpha} k_{3 \mu} \nonumber \\
&=& (n-2)! \left( F_{2}^{\alpha 0 \ldots p} k_4^{\mu} C^{(4)}_{0 \ldots p} + (-1)^{n-1} F_{2}^{\alpha \beta \mu_3 2 \ldots p} k_{4 \mu_3} C^{(4)}_{2 \ldots p}{}^{\mu \nu}  \eta_{\parallel \beta \nu} \right) k_{3 \alpha} k_{3 \mu} \nonumber \\
&=& (n-2)! (k_3 \cdot k_4) \left((k_2 \cdot k_3) C^{(2) 0 \ldots p} + E (k_3 \cdot C^{(2)} )^{1 \ldots p} \right) C^{(4)}_{0 \ldots p} \;,
\end{eqnarray}
where in the third line we have again used the fact that the index $\mu$ must lie parallel to the D-branes but $(k_3)_{\parallel}=0$.

\section{Deflection Angle for D-dimensional Schwarzschild Black Hole}\label{Ddimdeflection}

The metric for the D-dimensional Schwarzschild black hole is given by,
\begin{equation}
ds^2 = - \left(1-\frac{R_s^n}{r^n} \right) dt^2 + \left(1-\frac{R_s^n}{r^n} \right)^{-1} dr^2 + r^2 d \Omega_{n+1}^2 \;,
\end{equation}
where $n=D-3$ and $R_s$ is the Schwarzschild radius. Note that the Schwarzschild radius is related to the various constants of the stack of ($p=0$) D-branes by,

\begin{equation}
R_s^n = \frac{N \kappa_D T_{p=0}}{\Omega_{n+1}(n+1)} \;.
\end{equation}
We can express the deflection angle of a test probe in this background metric as,
\begin{equation}
\Phi = 2 \int_{r_0}^{\infty}  \text{d}r \; \frac{1}{r^2 \sqrt{\frac{1}{b^2} - \frac{1}{r^2} +\frac{1}{r^2} \left( \frac{R_s}{r} \right)^n}} - \pi \;, \label{eq:deflect1}
\end{equation}
where $r_0$ is the point of closest approach and $b$ is the impact parameter. The point of closest approach is obtained by solving,
\begin{equation}
\frac{1}{b^2} - \frac{1}{r_0^2} +\frac{1}{r_0^2} \left( \frac{R_s}{r_0} \right)^n = 0 \;. \label{eq:deflect2}
\end{equation}
Since $r_0 \gg R_s$, to first order we have $r_0 \approx b$. Note that if we write $u_0 = b/r_0$, then \eqref{eq:deflect2} becomes $1 - u_0^2 +u_0^{2+n}(R_s/b)^n = 0$. We can then solve this perturbatively and to second order the solution is given by $u_0 \approx 1 +c(R_s/b)^n$ where $c$ is found to be $1/2$ by substitution. In terms of $r_0$ this corresponds to letting $r_0 \approx b (1 + c (R_s/b)^n)$ with $c=-1/2$. Substituting \eqref{eq:deflect2} into \eqref{eq:deflect1} yields,
\begin{equation}
\Phi = 2 \int_{r_0}^{\infty}  \text{d}r \; \frac{1}{r^2 \sqrt{\frac{1}{r_0^2} -\frac{1}{r_0^2}\left( \frac{R_s}{r_0} \right)^n - \frac{1}{r^2} +\frac{1}{r^2} \left( \frac{R_s}{r} \right)^n}} - \pi \;.
\end{equation}
If we then perform the substitution $u=r_0/r$ we find that,
\begin{equation}
\Phi = 2 \int_{0}^{1}  \text{d}u \; \frac{1}{\sqrt{1 - \left( \frac{R_s}{r_0} \right)^n - u^2 + u^{2+n} \left( \frac{R_s}{r_0} \right)^n}} - \pi \;.
\end{equation}
By using the binomial expansion we find to second order in $(R_s/r_0)^n$,
\begin{eqnarray}
\Phi &=& 2 \int_{0}^{1}  \text{d}u \; \frac{1}{\sqrt{1 - u^2}} \left(1 + \frac{u^{2+n}-1}{1-u^2}  \left( \frac{R_s}{r_0} \right)^n \right)^{-1/2} - \pi \nonumber\\
& \approx & 2 \int_{0}^{1}  \text{d}u \; \frac{1}{\sqrt{1 - u^2}} + 2 \int_{0}^{1}  \text{d}u \; \frac{1-u^{2+n}}{2(1 - u^2)^{3/2}} \left( \frac{R_s}{r_0} \right)^n \nonumber \\
&& + 2 \int_{0}^{1}  \text{d}u \; \frac{3}{8} \frac{(1-u^{2+n})^2}{(1 - u^2)^{5/2}} \left( \frac{R_s}{r_0} \right)^{2n} - \pi \;. \label{eq:deflect3}
\end{eqnarray}
Each of the integrals over $u$ can be readily solved,
\begin{equation}
2 \int_{0}^{1}  \text{d}u \; \frac{1}{\sqrt{1 - u^2}} = \pi \;,
\end{equation}

\begin{equation}
2 \int_{0}^{1}  \text{d}u \; \frac{1-u^{2+n}}{2(1 - u^2)^{3/2}} = \sqrt{\pi} \frac{\Gamma{\left(\frac{3+n}{2}\right)}}{\Gamma{{\left(\frac{2+n}{2}\right)}}} \;,
\end{equation}

\begin{equation}
2 \int_{0}^{1}  \text{d}u \; \frac{3}{8} \frac{(1-u^{2+n})^2}{(1 - u^2)^{5/2}} = \frac{\sqrt{\pi}}{2} \left( \frac{\Gamma{\left(\frac{5+2n}{2}\right)}}{\Gamma{{\left( 1+n \right)}}}  - \frac{2\Gamma{\left(\frac{3+n}{2}\right)}}{\Gamma{{\left(\frac{n}{2}\right)}}} \right) \;.
\end{equation}
If we substitute these results into \eqref{eq:deflect3} we find,
\begin{equation}
\Phi = \sqrt{\pi} \frac{\Gamma{\left(\frac{3+n}{2}\right)}}{\Gamma{{\left(\frac{2+n}{2}\right)}}}  \left( \frac{R_s}{r_0} \right)^{n} + \frac{\sqrt{\pi}}{2} \left( \frac{\Gamma{\left(\frac{5+2n}{2}\right)}}{\Gamma{{\left( 1+n \right)}}}  - \frac{2\Gamma{\left(\frac{3+n}{2}\right)}}{\Gamma{{\left(\frac{n}{2}\right)}}} \right) \left( \frac{R_s}{r_0} \right)^{2n} + \ldots \label{eq:deflect4}
\end{equation}
We want to express this in terms of the impact parameter $b$. As we mentioned before we can perturbatively write $r_0$ in terms of the impact parameter as $r_0 \approx b ( 1 - 1/2 (R_s/b)^n )$. If we substitute this into \eqref{eq:deflect4}, keeping terms up to second order we find,
\begin{equation}
\Phi = \sqrt{\pi} \frac{\Gamma{\left(\frac{3+n}{2}\right)}}{\Gamma{{\left(\frac{2+n}{2}\right)}}}  \left( \frac{R_s}{b} \right)^{n} + \frac{\sqrt{\pi}}{2} \frac{\Gamma{\left(\frac{5+2n}{2}\right)}}{\Gamma{{\left( 1+n \right)}}} \left( \frac{R_s}{b} \right)^{2n} + \ldots \label{eq:deflect5}
\end{equation}

\providecommand{\href}[2]{#2}\begingroup\raggedright\endgroup


\begin{thebibliography}{10}

\bibitem{tHooft:1987vrq}
G.~'t~Hooft, ``{Graviton Dominance in Ultrahigh-Energy Scattering},''
\href{http://dx.doi.org/10.1016/0370-2693(87)90159-6}{{\em Phys. Lett.}
  {\bfseries B198} (1987) 61--63}.

\bibitem{Amati:1987wq}
D.~Amati, M.~Ciafaloni, and G.~Veneziano, ``{Superstring Collisions at
  Planckian Energies},''
\href{http://dx.doi.org/10.1016/0370-2693(87)90346-7}{{\em Phys. Lett.}
  {\bfseries B197} (1987) 81}.

\bibitem{Muzinich:1987in}
I.~J. Muzinich and M.~Soldate, ``{High-Energy Unitarity of Gravitation and
  Strings},''
\href{http://dx.doi.org/10.1103/PhysRevD.37.359}{{\em Phys. Rev.} {\bfseries
  D37} (1988) 359}.

\bibitem{Giddings:2006vu}
S.~B. Giddings, ``{Locality in quantum gravity and string theory},''
  \href{http://dx.doi.org/10.1103/PhysRevD.74.106006}{{\em Phys. Rev.}
  {\bfseries D74} (2006) 106006},
\href{http://arxiv.org/abs/hep-th/0604072}{{\ttfamily arXiv:hep-th/0604072}}.

\bibitem{D'Appollonio:2013hja}
G.~D'Appollonio, P.~Vecchia, R.~Russo, and G.~Veneziano, ``{Microscopic unitary
  description of tidal excitations in high-energy string-brane collisions},''
  \href{http://dx.doi.org/10.1007/JHEP11(2013)126}{{\em JHEP} {\bfseries 1311}
  (2013) 126},
\href{http://arxiv.org/abs/1310.1254}{{\ttfamily arXiv:1310.1254 [hep-th]}}.

\bibitem{Amati:1987uf}
D.~Amati, M.~Ciafaloni, and G.~Veneziano, ``{Classical and Quantum Gravity
  Effects from Planckian Energy Superstring Collisions},''
\href{http://dx.doi.org/10.1142/S0217751X88000710}{{\em Int. J. Mod. Phys.}
  {\bfseries A3} (1988) 1615--1661}.

\bibitem{Amati:1988tn}
D.~Amati, M.~Ciafaloni, and G.~Veneziano, ``{Can Space-Time Be Probed Below the
  String Size?},''
\href{http://dx.doi.org/10.1016/0370-2693(89)91366-X}{{\em Phys. Lett.}
  {\bfseries B216} (1989) 41}.

\bibitem{D'Appollonio:2010ae}
G.~D'Appollonio, P.~Di~Vecchia, R.~Russo, and G.~Veneziano, ``{High-energy
  string-brane scattering: Leading eikonal and beyond},''
  \href{http://dx.doi.org/10.1007/JHEP11(2010)100}{{\em JHEP} {\bfseries 1011}
  (2010) 100}, \href{http://arxiv.org/abs/1008.4773}{{\ttfamily arXiv:1008.4773
  [hep-th]}}.

\bibitem{Camanho:2014apa}
X.~O. Camanho, J.~D. Edelstein, J.~Maldacena, and A.~Zhiboedov, ``{Causality
  Constraints on Corrections to the Graviton Three-Point Coupling},''
  \href{http://dx.doi.org/10.1007/JHEP02(2016)020}{{\em JHEP} {\bfseries 02}
  (2016) 020},
\href{http://arxiv.org/abs/1407.5597}{{\ttfamily arXiv:1407.5597 [hep-th]}}.

\bibitem{DAppollonio:2015fly}
G.~D'Appollonio, P.~Di~Vecchia, R.~Russo, and G.~Veneziano, ``{Regge behavior
  saves String Theory from causality violations},''
  \href{http://dx.doi.org/10.1007/JHEP05(2015)144}{{\em JHEP} {\bfseries 05}
  (2015) 144},
\href{http://arxiv.org/abs/1502.01254}{{\ttfamily arXiv:1502.01254 [hep-th]}}.

\bibitem{Giddings:2010pp}
S.~B. Giddings, M.~Schmidt-Sommerfeld, and J.~R. Andersen, ``{High energy
  scattering in gravity and supergravity},''
  \href{http://dx.doi.org/10.1103/PhysRevD.82.104022}{{\em Phys. Rev.}
  {\bfseries D82} (2010) 104022},
\href{http://arxiv.org/abs/1005.5408}{{\ttfamily arXiv:1005.5408 [hep-th]}}.

\bibitem{Akhoury:2013yua}
R.~Akhoury, R.~Saotome, and G.~Sterman, ``{High Energy Scattering in
  Perturbative Quantum Gravity at Next to Leading Power},''
\href{http://arxiv.org/abs/1308.5204}{{\ttfamily arXiv:1308.5204 [hep-th]}}.

\bibitem{Melville:2013qca}
S.~Melville, S.~G. Naculich, H.~J. Schnitzer, and C.~D. White, ``{Wilson line
  approach to gravity in the high energy limit},''
  \href{http://dx.doi.org/10.1103/PhysRevD.89.025009}{{\em Phys. Rev.}
  {\bfseries D89} no.~2, (2014) 025009},
\href{http://arxiv.org/abs/1306.6019}{{\ttfamily arXiv:1306.6019 [hep-th]}}.

\bibitem{Bjerrum-Bohr:2014zsa}
N.~E.~J. Bjerrum-Bohr, J.~F. Donoghue, B.~R. Holstein, L.~Planté, and
  P.~Vanhove, ``{Bending of Light in Quantum Gravity},''
  \href{http://dx.doi.org/10.1103/PhysRevLett.114.061301}{{\em Phys. Rev.
  Lett.} {\bfseries 114} no.~6, (2015) 061301},
\href{http://arxiv.org/abs/1410.7590}{{\ttfamily arXiv:1410.7590 [hep-th]}}.

\bibitem{Bjerrum-Bohr:2016hpa}
N.~E.~J. Bjerrum-Bohr, J.~F. Donoghue, B.~R. Holstein, L.~Plante, and
  P.~Vanhove, ``{Light-like Scattering in Quantum Gravity},''
  \href{http://dx.doi.org/10.1007/JHEP11(2016)117}{{\em JHEP} {\bfseries 11}
  (2016) 117},
\href{http://arxiv.org/abs/1609.07477}{{\ttfamily arXiv:1609.07477 [hep-th]}}.

\bibitem{Luna:2016idw}
A.~Luna, S.~Melville, S.~G. Naculich, and C.~D. White, ``{Next-to-soft
  corrections to high energy scattering in QCD and gravity},''
  \href{http://dx.doi.org/10.1007/JHEP01(2017)052}{{\em JHEP} {\bfseries 01}
  (2017) 052},
\href{http://arxiv.org/abs/1611.02172}{{\ttfamily arXiv:1611.02172 [hep-th]}}.

\bibitem{Amati:1990xe}
D.~Amati, M.~Ciafaloni, and G.~Veneziano, ``{Higher order gravitational
  deflection and soft bremsstrahlung in Planckian energy superstring
  collisions},''
\href{http://dx.doi.org/10.1016/0550-3213(90)90375-N}{{\em Nucl. Phys.}
  {\bfseries B347} (1990) 550--580}.

\bibitem{Buonanno:1998gg}
A.~Buonanno and T.~Damour, ``{Effective one-body approach to general
  relativistic two-body dynamics},''
  \href{http://dx.doi.org/10.1103/PhysRevD.59.084006}{{\em Phys. Rev.}
  {\bfseries D59} (1999) 084006},
\href{http://arxiv.org/abs/gr-qc/9811091}{{\ttfamily arXiv:gr-qc/9811091
  [gr-qc]}}.

\bibitem{Damour:2016gwp}
T.~Damour, ``{Gravitational scattering, post-Minkowskian approximation and
  Effective One-Body theory},''
  \href{http://dx.doi.org/10.1103/PhysRevD.94.104015}{{\em Phys. Rev.}
  {\bfseries D94} no.~10, (2016) 104015},
\href{http://arxiv.org/abs/1609.00354}{{\ttfamily arXiv:1609.00354 [gr-qc]}}.

\bibitem{Damour:2017zjx}
T.~Damour, ``{High-energy gravitational scattering and the general relativistic
  two-body problem},'' \href{http://dx.doi.org/10.1103/PhysRevD.97.044038}{{\em
  Phys. Rev.} {\bfseries D97} no.~4, (2018) 044038},
\href{http://arxiv.org/abs/1710.10599}{{\ttfamily arXiv:1710.10599 [gr-qc]}}.

\bibitem{Klebanov:1995ni}
I.~R. Klebanov and L.~Thorlacius, ``{The Size of p-branes},''
  \href{http://dx.doi.org/10.1016/0370-2693(95)01576-0}{{\em Phys.Lett.}
  {\bfseries B371} (1996) 51--56},
  \href{http://arxiv.org/abs/hep-th/9510200}{{\ttfamily arXiv:hep-th/9510200
  [hep-th]}}.

\bibitem{Garousi:1996ad}
M.~R. Garousi and R.~C. Myers, ``{Superstring Scattering from D-Branes},''
  \href{http://dx.doi.org/10.1016/0550-3213(96)00316-1}{{\em Nucl. Phys.}
  {\bfseries B475} (1996) 193--224},
\href{http://arxiv.org/abs/hep-th/9603194}{{\ttfamily arXiv:hep-th/9603194}}.

\bibitem{Hashimoto:1996bf}
A.~Hashimoto and I.~R. Klebanov, ``{Scattering of strings from D-branes},''
  {\em Nucl.Phys.Proc.Suppl.} {\bfseries 55B} (1997) 118--133,
  \href{http://arxiv.org/abs/hep-th/9611214}{{\ttfamily arXiv:hep-th/9611214
  [hep-th]}}.

\bibitem{Bakhtiarizadeh:2013zia}
H.~R. Bakhtiarizadeh and M.~R. Garousi, ``{Sphere-level Ramond-Ramond couplings
  in Ramond-Neveu-Schwarz formalism},''
  \href{http://dx.doi.org/10.1016/j.nuclphysb.2014.05.002}{{\em Nucl. Phys.}
  {\bfseries B884} (2014) 408--437},
\href{http://arxiv.org/abs/1312.4703}{{\ttfamily arXiv:1312.4703 [hep-th]}}.

\bibitem{green1988superstring}
M.~Green, M.~Green, J.~Schwarz, and E.~Witten, {\em Superstring Theory: Volume
  1, Introduction}.
\newblock Cambridge Monographs on Mathematical Physics. Cambridge University
  Press, 1988.

\bibitem{Garousi:2012yr}
M.~R. Garousi, ``{T-duality of the Riemann curvature corrections to
  supergravity},'' \href{http://dx.doi.org/10.1016/j.physletb.2012.12.012}{{\em
  Phys. Lett.} {\bfseries B718} (2013) 1481--1488},
\href{http://arxiv.org/abs/1208.4459}{{\ttfamily arXiv:1208.4459 [hep-th]}}.

\bibitem{Duff:1994an}
M.~Duff, R.~R. Khuri, and J.~Lu, ``{String solitons},''
  \href{http://dx.doi.org/10.1016/0370-1573(95)00002-X}{{\em Phys.Rept.}
  {\bfseries 259} (1995) 213--326},
\href{http://arxiv.org/abs/hep-th/9412184}{{\ttfamily arXiv:hep-th/9412184
  [hep-th]}}.

\bibitem{BjerrumBohr:2002ks}
N.~E.~J. Bjerrum-Bohr, J.~F. Donoghue, and B.~R. Holstein, ``{Quantum
  corrections to the Schwarzschild and Kerr metrics},''
  \href{http://dx.doi.org/10.1103/PhysRevD.68.084005,
  10.1103/PhysRevD.71.069904}{{\em Phys. Rev.} {\bfseries D68} (2003) 084005},
  \href{http://arxiv.org/abs/hep-th/0211071}{{\ttfamily arXiv:hep-th/0211071
  [hep-th]}}.
[Erratum: Phys. Rev.D71,069904(2005)].

\bibitem{Bjerrum-Bohr:2018xdl}
N.~E.~J. Bjerrum-Bohr, P.~H. Damgaard, G.~Festuccia, L.~Planté, and
  P.~Vanhove, ``{General Relativity from Scattering Amplitudes},''
\href{http://arxiv.org/abs/1806.04920}{{\ttfamily arXiv:1806.04920 [hep-th]}}.

\bibitem{Emparan:2009at}
R.~Emparan, T.~Harmark, V.~Niarchos, and N.~A. Obers, ``{Essentials of
  Blackfold Dynamics},'' \href{http://dx.doi.org/10.1007/JHEP03(2010)063}{{\em
  JHEP} {\bfseries 03} (2010) 063},
\href{http://arxiv.org/abs/0910.1601}{{\ttfamily arXiv:0910.1601 [hep-th]}}.

\bibitem{Duff:1974xx}
M.~J. Duff, ``{Quantum tree graphs and the Schwarzschild solution},''
\href{http://dx.doi.org/10.1103/PhysRevD.7.2317}{{\em Phys. Rev.} {\bfseries
  D7} (1973) 2317--2326}.

\bibitem{Bertolini:2000jy}
M.~Bertolini {\em et~al.}, ``{Is a classical description of stable non-BPS
  D-branes possible?},''
  \href{http://dx.doi.org/10.1016/S0550-3213(00)00558-7}{{\em Nucl. Phys.}
  {\bfseries B590} (2000) 471--503},
\href{http://arxiv.org/abs/hep-th/0007097}{{\ttfamily arXiv:hep-th/0007097}}.

\bibitem{Ademollo:1989ag}
M.~Ademollo, A.~Bellini, and M.~Ciafaloni, ``{Superstring Regge amplitudes and
  emission vertices},''
\href{http://dx.doi.org/10.1016/0370-2693(89)91609-2}{{\em Phys.Lett.}
  {\bfseries B223} (1989) 318--324}.

\bibitem{Ademollo:1990sd}
M.~Ademollo, A.~Bellini, and M.~Ciafaloni, ``{Superstring Regge amplitudes and
  graviton radiation at planckian energies},''
\href{http://dx.doi.org/10.1016/0550-3213(90)90626-O}{{\em Nucl.Phys.}
  {\bfseries B338} (1990) 114--142}.

\bibitem{Brower:2006ea}
R.~C. Brower, J.~Polchinski, M.~J. Strassler, and C.-I. Tan, ``{The Pomeron and
  gauge/string duality},''
  \href{http://dx.doi.org/10.1088/1126-6708/2007/12/005}{{\em JHEP} {\bfseries
  0712} (2007) 005},
\href{http://arxiv.org/abs/hep-th/0603115}{{\ttfamily arXiv:hep-th/0603115
  [hep-th]}}.

\bibitem{Bini:2017xzy}
D.~Bini and T.~Damour, ``{Gravitational spin-orbit coupling in binary systems,
  post-Minkowskian approximation and effective one-body theory},''
  \href{http://dx.doi.org/10.1103/PhysRevD.96.104038}{{\em Phys. Rev.}
  {\bfseries D96} no.~10, (2017) 104038},
\href{http://arxiv.org/abs/1709.00590}{{\ttfamily arXiv:1709.00590 [gr-qc]}}.

\bibitem{Vines:2017hyw}
J.~Vines, ``{Scattering of two spinning black holes in post-Minkowskian
  gravity, to all orders in spin, and effective-one-body mappings},''
  \href{http://dx.doi.org/10.1088/1361-6382/aaa3a8}{{\em Class. Quant. Grav.}
  {\bfseries 35} no.~8, (2018) 084002},
\href{http://arxiv.org/abs/1709.06016}{{\ttfamily arXiv:1709.06016 [gr-qc]}}.

\end{thebibliography}
\end{document}